\input cp-aa.inp
\catcode`@=\active   
\def\dd{{\rm d}}
\def\note #1]{{\bf #1]}}
\def\etal{~et~al.}
\def\ac{\alpha}
\def\ist{\,=\,}
\def\figdir{./}

\raggedbottom
%
%
\MAINTITLE={ Amplitudes of stochastically excited oscillations in main-sequence stars}
\SUBTITLE={ ????? }
\AUTHOR={ G. Houdek@{1,2}, N.J. Balmforth@3, 
          J. Christensen-Dalsgaard@4, and D.O. Gough@{1,5}}
\OFFPRINTS={ G. Houdek }
\INSTITUTE={@1 Institute of Astronomy, University of Cambridge, 
               Cambridge CB3 0HA, England
            @2 Institut f\"ur Astronomie, Universit\"at Wien,
               A-1180 Wien, Austria; e-mail: hg\at ast.cam.ac.uk
            @3 Instituto di Cosmogeofisica, Corso Fiume 4, Torino 10133, Italy;
               e-mail: njb\at hank.ucsd.edu
            @4 Teoretisk Astrofysik Center, Danmarks Grundforskningsfond, and
               Institut for Fysik og Astronomi,\hfil\break
              \hbox{\phantom{@3}Aarhus Universitet, DK-8000 Aarhus C, Denmark;
               e-mail: jcd\at obs.aau.dk}
            @5 Department of Applied Mathematics and Theoretical Physics,
               University of Cambridge,\hfil\break
               \hbox{\phantom{@5}Cambridge CB3 9EW, England; 
               e-mail: douglas\at ast.cam.ac.uk}
           }
\DATE={ Received month day, 1999, accepted month day, 1999 }
\ABSTRACT={ We present estimates of the amplitudes of intrinsically stable
            stochastically excited radial oscillations in stars near the main
            sequence. The amplitudes are determined by the balance between 
            acoustical energy generation by turbulent convection (the
            Lighthill mechanism) and linear damping. Convection is treated 
            with a time-dependent, nonlocal, mixing-length model, which 
            includes both convective heat flux and turbulent pressure in both 
            the equilibrium model and the pulsations. Velocity and luminosity 
            amplitudes are computed for stars with masses between 
            $0.9\,M_\odot$ and $2.0\,M_\odot$ in the vicinity of the main 
            sequence, for various metallicities and convection parameters. 
            As in previous studies, the amplitudes are found to increase with 
            stellar mass, and therefore with luminosity.
            Amongst those stars that are pulsationally stable, the largest 
            amplitudes are predicted for a $1.6\,M_\odot$ model of spectral 
            type F$2$; the values are approximately $15$ times larger 
            than those measured in the Sun.
          }
\KEYWORDS={ main-sequence stars -- convection -- pulsational stability  }
\THESAURUS={ 02.03.3; 02.20.1; 06.15.1; 08.15.1; 08.22.2 }
\maketitle
\AUTHORRUNNINGHEAD{Houdek, Balmforth, Christensen-Dalsgaard and Gough}
\MAINTITLERUNNINGHEAD{Amplitudes of stochastically excited oscillations}
\titlea{Introduction}
The stability of solar-like p modes depends mainly on the
interaction of the oscillations with radiation and convection
in the outer envelope. The most plausible explanations for the occurrence
of such oscillations are either intrinsic thermal overstability or
stochastic excitation of stable modes by turbulent convection.
Whatever the mechanism, the energy flow from radiation and convection
into and out of the p modes takes place very near the surface
(e.g. Goode, Gough \& Kosovichev 1992).
Sun-like stars possess surface convection zones, and it is in 
these zones, where the energy is transported principally by the 
turbulence, that most of the driving takes place.
Mode stability is governed not only by the perturbations in the radiative 
fluxes (via the $\kappa$-mechanism) but also by the perturbations in the 
turbulent fluxes (heat and momentum). The study of mode stability therefore 
demands a theory for convection that includes the interaction of the 
turbulent velocity field with the pulsation.

Thermal overstability of pulsations arises from an exchange of energy 
between the oscillatory motion of the stellar matter, the turbulence 
and the radiation field.  Such overstability has been suggested as a 
possible mechanism for the excitation of solar oscillations by 
Ulrich (1970a) and Antia, Chitre \& Gough (1988). If solar p modes 
were indeed overstable, some nonlinear mechanism
must limit their amplitudes to the values that are observed.
Nonlinear coupling to other, stable modes was considered by
Kumar \& Goldreich (1989), who estimated that the energy drain through
three-mode coupling would occur at a rate too low to extract the energy gained
from the overstability at the appropriate amplitude. Similar estimates of
nonlinear self limiting are also too weak. The saturation of mode amplitudes 
at the observed levels therefore remains a mystery if overstability provides 
the origin of solar pulsations.

The problem of identifying a saturation mechanism does not arise if the modes
are intrinsically stable. Such modes can be stochastically excited by the
turbulent convection. The process can be regarded as multipole acoustical 
radiation (e.g. Unno 1964). For solar-like stars, the acoustic
noise generated by convection in the star's resonant cavity 
may be manifest as an ensemble of p modes over a wide band in 
frequency (Goldreich \& Keeley 1977b). The amplitudes are determined by 
the balance between the excitation and damping, and are expected to be 
rather low. The turbulent-excitation model predicts not 
only the right order of magnitude for the p-mode amplitudes (Gough 1980), 
but it also explains the observation that millions of modes are excited 
simultaneously. 
Moreover, recent observations (Toutain \& Fr\"ohlich 1992,
Goode \& Strous 1996; Chaplin\etal\ 1998) also support a stochastic origin.
This second explanation seems therefore to be the more likely, and is 
the one that we shall adopt here.

To date, Christensen-Dalsgaard \& 
Frandsen (1983b) have made the only predictions of amplitudes of 
solar-like oscillations in other stars.
They obtained amplitudes of modes
by postulating equipartition between the energy of an
oscillation mode and the kinetic energy in one convective eddy having the same
turnover time as the period of the oscillation. This simple formula for 
excitation was proposed by Goldreich \& Keeley (1977b), who used it to 
estimate amplitudes for the solar case, assuming damping rates determined
solely by a scalar turbulent viscosity (Goldreich \& Keeley 1977a).
In the calculations of Christensen-Dalsgaard \& Frandsen (1983b), however,
radial eigenfunctions of model envelopes were computed by solving the 
equations of linear nonadiabatic oscillation, although they also neglected 
turbulent pressure and set the Lagrangian perturbation to the convective 
heat flux to zero. They found velocity and luminosity amplitudes 
to increase with age, and with increasing mass along the main sequence. 

Balmforth (1992a) improved the calculation by introducing Gough's (1976, 1977)
nonlocal, time-dependent mixing-length model for convection, using the 
Eddington approximation to radiative transfer for both the equilibrium 
structure and the pulsations. He calculated damping rates for the solar 
case and found all modes to be stable. Here we continue Balmforth's 
investigation and study the oscillations of main-sequence stars, delimiting 
the region in the HR diagram for stars with stable modes. Preliminary results 
of the calculations have been presented by Houdek\etal\ (1995).

According to Libbrecht\etal\ (1986) the observed oscillation properties of 
low-degree modes depend little on the value of $l$. This is to be expected
because the excitation and damping mechanisms are significant only very close
to the surface, where the vertical scale is much less than the horizontal 
scale of oscillations and when $l$ is low the modal inertia is quite 
insensitive to degree $l$. It is therefore adequate to simplify the 
calculations, by analysing only radial modes of oscillation. The results 
are applicable to all modes of moderately low degree.

\titlea{Observational projects}    
Observations of oscillation properties of stars other than the Sun
provide important information for testing the theory of stellar evolution.
A critical problem with the detection of oscillations in solar-type stars, 
however, is their very small amplitudes, of the order of $1$\,m\,s$^{-1}$ 
or less. For the Sun the observed velocity variations in disc-integrated light
have values $\la 20$\,cm\,s$^{-1}$ (e.g. Grec\etal\ 1983; Libbrecht \& Woodard
1991; Chaplin\etal\ 1998). To detect similar variations in distant stars is 
therefore a challenging task, requiring observations to be made with the utmost
precision. 

Three observing techniques for detecting such oscillations have been
developed so far. The first is to search for periodic Doppler shifts 
of spectral lines (e.g. Kennelley 1995). However, the most successful result 
by this method has been the determination of only an upper bound to 
oscillation amplitudes in some of the brightest stars
(e.g. Brown \& Gilliland 1990; Brown\etal\ 1991; Mosser\etal\ 1998).
The second method is to look for periodic brightness fluctuations using
photometry. 
When used with area detectors such as CCDs, this method has a clear advantage
over spectroscopic techniques because it permits one to observe large 
ensembles of stars simultaneously 
(e.g. Gilliland 1995). 
Using differential CCD photometry
with seven $4$m-class telescopes, Gilliland\etal\ (1993) obtained upper
limits of luminosity amplitudes of possible oscillations in twelve
stars in ${\rm M}67$. The third method, introduced by   
Kjeldsen \& Bedding (1995), measures temperature fluctuations induced
by stellar oscillations via their effects on the equivalent width of the
Balmer lines (Bedding\etal\ 1996). This %
technique has yielded a possible detection of solar-like oscillations in 
the sub-giant $\eta$~Boo (Kjeldsen et al.\ 1995).  Although it is currently 
restricted to observing isolated stars, the equivalent-width method is 
insensitive to atmospheric scintillation, and attains a substantially better 
signal-to-noise ratio than do the other two ground-based methods. 

The limitations of ground-based observational techniques in asteroseismology 
have been addressed by Frandsen (1992) and Gilliland (1995), both of whom
argued that seismology can be applied to distant solar-type stars only by 
observing them from space. The elimination of atmospheric noise and the 
possibility of obtaining long continuous data sequences will provide 
information of much higher quality than from any ground-based method. 
Several asteroseismological space projects are in preparation, such as
the French project COROT\fonote{COnvection and ROTation} (Catala\etal\ 1995),
the Danish project MONS\fonote{Measuring Oscillations in Nearby Stars} 
(Kjeldsen \& Bedding 1998) and the Canadian project MOST\fonote{
Microvariability \& Oscillations of STars} (Matthews 1998).

When preparing an observing campaign, it is helpful in the selection of
target stars to have a good prediction of the amplitude of the signal 
one will be trying to observe. 
Measurements of mode lifetimes (damping rates) and the variation of 
oscillation amplitudes, which depend, in part, on stellar parameters,
provide invaluable insights into the mechanisms that excite solar-like 
oscillations. It is hoped that future observations will provide these 
crucial measurements. 
The aim of this paper is to provide a systematic
survey of the oscillation properties in view of these upcoming
observational projects.

\titlea{Time-dependent convection}
In order to describe the turbulent fluxes in a time-dependent
envelope, various phenomenological mixing-length models have been
proposed (for a detailed discussion see Balmforth 1992a; Houdek 1996).
In envelopes that do not pulsate, the various guises of local mixing-length
method are essentially similar, once their intrinsic parameters
have been calibrated (their formulations may be interpreted as providing
interpolation formulae between the two limits of efficient and inefficient 
convection; Gough \& Weiss 1976). 
This is not so of formulations of time-dependent mixing-length prescriptions,
in which the details of the phenomenological model influence the
predictions of stability (Balmforth 1992a).
Moreover, local theories are plagued by some fundamental inconsistencies, 
which we shall describe presently. For these reasons, we 
prefer to use a nonlocal version (Gough 1976) which is summarized below.

\titleb{Local mixing-length model}
Local mixing-length models
(e.g. B\"ohm-Vitense 1958; Ulrich 1970b; Gough 1977) still provide
the almost universal method for computing the stratification of convection 
zones in stellar models.
One of the major drawbacks of a local approach is the assumption that the 
characteristic length scale $\ell$ must be shorter than any scale
associated with the structure of the star. This condition is certainly
violated in solar-like stars and in red giants, where calibrated evolution 
calculations yield a typical value for the mixing-length parameter 
$\ac=\ell/H_p$ that is of order unity, where $H_p$ is the pressure 
scale height. 
(Since $\ac$ is normally thought to be essentially invariable, it follows
that the condition must be thought to be violated in all stars.) 
This implies that fluid properties vary significantly over the extent of 
a convective element; the superadiabatic gradient can vary on a scale 
much shorter than $\ell$.

In many computations of stellar envelope models, the turbulent 
pressure $p_{\rm t}$ associated with the Reynolds stresses has been 
ignored. However, several investigations (Baker \& Gough 1979; 
Rosenthal\etal\ 1995; Canuto \& Christensen-Dalsgaard 1998) suggest that 
the momentum flux provides a substantial fraction of the hydrostatic support 
in the equilibrium model, and should therefore not be neglected.
In a local model,
inclusion of the momentum flux $p_{\rm t}=\rho w^2$ ($\rho$ being the mean 
density and $w$ the rms vertical velocity of the convective elements) leads 
to singular points in the equation for the turbulent pressure gradient at the 
edges of the convective regions (e.g. Gough 1976). 
This issue demands careful consideration, although it can be circumvented by
adopting an additional approximation to reduce the order of the equations of 
hydrostatic support, which are then not strictly consistent with the 
formulation of the theory (Henyey, Vardya \& Bodenheimer 1965; 
Baker \& Gough 1979).

In a time-dependent description, the details of the phenomenological
model become important to the linear stability calculation;
the nonlocal model that we employ in our calculations is one
that is based on Gough's (1965, 1977) local mixing-length model.
In that formulation, the turbulent eddies that support the heat and momentum
fluxes evolve in a pulsating environment. Explicit consideration
is given to the phase of pulsation at the birth of
each transitorily coherent eddy, and to how the eddy adjusts to
the temporally varying environment. Other mixing-length models
(e.g. Unno 1967, Xiong 1989) emphasize other aspects of the dynamics, which
presumably influence differently the pulsational stability (Balmforth 1992a).

Another major drawback of local theory is that it fails to treat the convective
dynamics across extended eddies in a physically plausible fashion:
in deeper parts of the convection zone, where 
the stratification is almost adiabatic, convective heat transport is very 
efficacious; radiative diffusion is unimportant, and the perturbation 
of the heat flux is dominated by advection of temperature fluctuations.
In this limit, the temperature fluctuation can be described approximately by
a diffusion equation in which the diffusivity is imaginary
(Baker \& Gough 1979; Gonczi \& Osaki 1980). 
Rapid spatial oscillations of the eigenfunctions result.
The problem this introduces can at best be thought of as one of numerical
resolution, which is particularly severe in layers where the stratification 
is very close to being adiabatic.
But worse, it signifies a complete failure of the model, since the wavelength 
of the spatial oscillation is very much smaller than the mixing length
(which is supposedly the smallest lengthscale permitted by the model).

\titleb{Nonlocal mixing-length model}
The obvious drawbacks of a local formulation of the mixing-length approach
can be removed by an appropriate nonlocal generalization. 
Account can be taken of the finite size of a convective eddy by averaging 
spatially the representative value of a physical variable throughout 
the eddy. 
Spiegel (1963), for example,  proposed a nonlocal description based on the 
concept of an eddy phase space, and derived an equation for the 
convective flux
which is similar to a radiative-transfer equation. The solution of this 
transfer equation yields an integral expression that converts the usual 
ordinary differential equations describing stellar models into 
integro-differential equations. The solution to the transfer equation can be 
approximated by taking moments, closing the hierarchy at second order 
with the 
Eddington approximation (Gough 1976). In this approximation, the nonlocal 
generalization of the mixing-length formulation is governed by three 
second-order differential equations for the spatially
averaged convective fluxes $F_{\rm c}$ and $p_{\rm t}$ and for the
average superadiabatic temperature gradient experienced by an eddy.

The procedure introduces two more parameters, $a$ and $b$, 
which characterize respectively the spatial coherence of the ensemble of 
eddies contributing to the total heat and momentum fluxes, and the extent over
which the turbulent eddies experience an average of the local stratification.
Theory suggests approximate values for these parameters, but it is
arguably better to treat them as free. Roughly speaking, 
the parameters control the degree of ``nonlocality'' of convection;
low values imply highly nonlocal solutions, and in the limit
$a,b\rightarrow\infty$ the system of equations reduces to the 
local formulation (except near the boundaries of the convection zone, where
the equations are singular).
Balmforth (1992a) investigated the effect of the parameters $a$ and $b$ on
the turbulent fluxes in the solar case, and 
Tooth \& Gough (1989) tried to calibrate $a$ and $b$ by comparing theory
with laboratory experiments.

\titlea{Model computations}
The basic model calculations reported in this paper are as described by 
Balmforth (1992a). In particular, we incorporate turbulent pressure in the 
equilibrium model envelope. The computation proceeds by iteration, from
a trial solution obtained by integrating 
inwards from an optical depth of $\tau = 10^{-4}$ and ending at a radius 
fraction $0.2$, using local mixing-length theory and the diffusion 
approximation to radiative transfer; 
the approximation to the turbulent pressure used by Baker \& Gough (1979)
is adopted to obviate singular points at the edges of the convection zones.
The entire envelope is then re-integrated using the equations appropriate to 
the nonlocal mixing-length theory, using the Eddington approximation to 
radiative transfer (Unno \& Spiegel 1966). The atmosphere is treated as being 
grey, and is assumed to be plane parallel. The temperature gradient is 
corrected by using a $\tau$-dependent varying Eddington factor 
(Auer \& Mihalas 1970) derived from model C of 
Vernazza, Avrett \& Loeser (1981) in the manner of Balmforth (1992a).
Opacities were obtained from the latest OPAL tables (Iglesias \& Rogers 1996),
supplemented at low temperature by tables from Kurucz (1991). 
Interpolation in these tables was carried out using birational splines
(Houdek \& Rogl 1996). 
The equation of state included a detailed treatment of the ionization
of C, N, and O, and a treatment of the first ionization of the next seven
most abundant elements (Christensen-Dalsgaard 1982), as well as 
`pressure ionization' by the method of Eggleton, Faulkner \& Flannery (1973); 
electrons were treated with relativistic Fermi-Dirac statistics.
In the pulsation model the boundary conditions used are essentially those of
Baker \& Kippenhahn (1965), but supplemented by appropriate conditions on the
variables of the nonlocal mixing-length theory (Balmforth 1992a). The outer 
boundary conditions were applied at the temperature minimum, the mechanical 
condition being consistent with a perfectly reflecting surface;
at the base of the envelope, conditions of adiabaticity and vanishing 
displacement were imposed.
 
The linearized pulsation equations were solved with a second-order accuracy
Newton-Raphson-Kantorovich algorithm (Baker, Moore \& Spiegel 1971; Gough,
Spiegel \& Toomre 1974). 
With this algorithm the eigenfunctions and eigenvalues can be computed 
simultaneously; however, one has to provide a proper trial solution. This 
can be obtained by solving first the adiabatic pulsation equations and then 
applying a quasi-adiabatic approximation\fonote{Quasi-adiabatic 
approximations adopt adiabatic eigenfunctions for evaluating the thermal
variables for computing mode stability.} to complete the nonadiabatic system.
The detailed equations describing the equilibrium and pulsation model have 
been discussed by Balmforth (1992a) and Houdek (1996).

\titlea{Damping rates}
Were solar p modes to be genuinely linear and stable, their power spectrum
could be described in terms of an ensemble of intrinsically damped, 
stochastically driven, simple-harmonic oscillators (Batchelor 1956; 
Christensen-Dalsgaard, Gough \& Libbrecht 1989) provided the background
equilibrium state of the star is independent of time; if we further assume
that mode phase fluctuations contribute negligibly to the width of the
spectral lines, the intrinsic damping rates of
the modes could then be determined observationally from measurements of the 
pulsation linewidths. 
The linewidths are obtained, to a first approximation, by
fitting Lorentzian profile functions to the spectral peaks of the observed 
power spectrum. Higher approximations demand a more detailed description of the 
excitation and damping (cf. Jefferies\etal\ 1988; Gabriel 1998). 
Continuous observations over many periods are required. Observations from 
GONG (Harvey\etal\ 1996), and from the SOI-MDI 
(Scherrer, Hoeksema \& Bush 1991), VIRGO (Fr\"ohlich\etal\ 1995) and GOLF 
(Gabriel\etal\ 1991) instruments aboard the SOHO spacecraft have already 
provided such high-quality data for the Sun.

It should be noted that the observed power spectrum of solar pulsations  
is complicated by the beating of closely spaced modes 
(e.g. Christensen-Dalsgaard \& Gough 1982; Hill\etal\ 1996)
and by amplitude variations and phase wandering resulting from the interaction
with the turbulent convection. These effects might also be responsible for
the observed asymmetries in the p-mode line profiles. The study of how to 
disentangle the manifestations of these phenomena from manifestations of 
the intrinsic mode parameters is only in its infancy 
(Chang, Gough \& Sekii 1997; Roxburgh \& Vorontsov 1997; Gabriel 1998;
Nigam\etal\ 1998; Rast \& Bogdan 1998). 
It is important for analysing not only the oscillations of the Sun, but the
oscillations of any star with a rich spectrum of frequencies.

\vskip -3truemm
\titleb{Processes contributing to intrinsic linewidths}

\begfigside 6.20cm 10.70cm
\figure{1}
{Physical processes contributing to the linear damping rate $\eta$.
They can be associated with the effects arising from the
momentum balance ($\eta_{\rm dyn}$) and from the thermal
energy balance ($\eta_{\rm g}$). The contributions $\eta_{\rm scatt}$
and $\eta_{\rm leak}$ are in parentheses because they have not been
taken into account in the computations reported in this paper.
The influence of Reynolds stresses on solar modes, 
contributing to $\eta_{\rm t}$, has been treated by 
Goldreich \& Keeley (1977a) in the manner of a time-independent
scalar turbulent viscosity.
The width of the line in the Fourier power spectrum of the oscillations
is influenced also by nonlinearities, both those coupling a mode to others
(Kumar \& Goldreich, 1989) and those intrinsic to the mode itself.
}
\includegraphics{\figdir/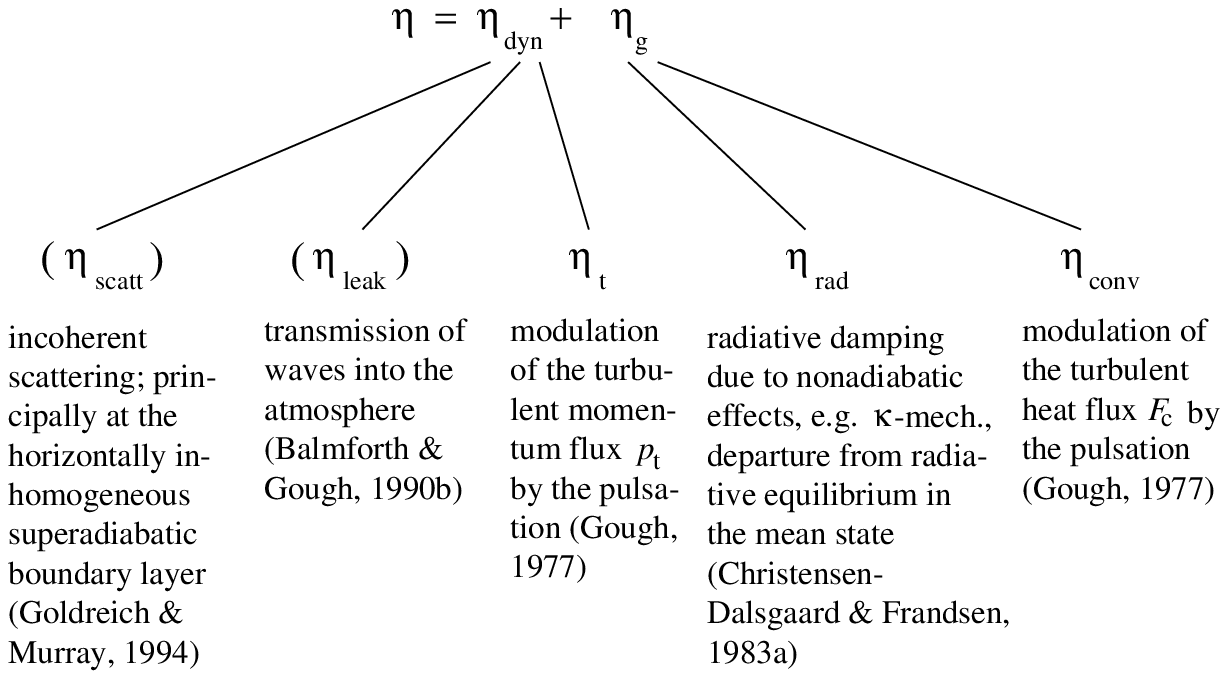}              
\endfig

Basically, the damping of stellar oscillations
arises from two sources: processes influencing the momentum balance, 
and processes influencing the thermal energy equation. Each of these 
contributions can be divided further according to their physical origin, as
illustrated in Fig.~1. A detailed discussion of the processes has been given 
by Houdek (1996). Here we limit the discussion to those that are modelled in 
our computations.

Nonadiabatic radiative processes can contribute to both the driving and the
damping of the pulsations. In solar-type stars the zones of ionization lie 
well inside the regions of efficacious convection, and the conventional 
$\kappa$-mechanism provides only a relatively small contribution to the driving.
Radiative damping in the atmosphere is not necessarily small, and requires a
more accurate treatment of radiative transfer than the diffusion approximation.
Christensen-Dalsgaard \& Frandsen (1983a) have shown that the use of the grey 
Eddington approximation, when applied correctly, does not introduce too
large an error in the calculation of the damping rates. Furthermore, they
have demonstrated that for stability calculations departures from radiative 
equilibrium in the mean state must not be neglected: in the upper boundary
layer of the convection zone, where there is a transition from convective to 
radiative energy transport, radiative equilibrium is no longer maintained. 
Thus the mean intensity $J$ is not equal to the Planck function $B$.
In particular, by perturbing the equations describing the radiation field in
the Eddington approximation, one obtains 
(Christensen-Dalsgaard \& Frandsen 1983a)
$$
  \delta({1\over\rho}{\rm div}\,{\mib F_{\rm r}})\ist
  4\pi\kappa[\delta B-\delta J+{\delta\kappa\over\kappa}(B_0-J_0)]
  \; ,\eqno\autnum
$$
where $\rho$ and $\kappa$ denote the density and opacity, respectively, 
the operator $\delta$ denotes a Lagrangian perturbation,
and the subscript 0 denotes an equilibrium quantity. 
The last term in Eq.\ (1) describes the departure from radiative 
equilibrium in the mean state; 
it is not everywhere small, yet it has been 
ignored in most stability calculations so far.

Vibrational stability is influenced further by the exchange of energy
between the pulsation and the turbulent velocity field. The exchange arises
either via the pulsationally perturbed convective heat flux, or directly 
through dynamical effects of the fluctuating Reynolds stresses.
In fact, it is the modulation of the turbulent fluxes by the pulsations that 
seems to be the predominant mechanism responsible for the driving and damping 
of solar-type acoustic modes.

Nonadiabatic processes attributed to the modulation of the convective 
heat flux by the pulsation are accounted for by the contribution 
$\eta_{\rm conv}$ to the total damping rates (see Fig.~1). This
contribution is related to the way that convection modulates large-scale
temperature perturbations induced by the pulsations, which influences 
pulsational stability substantially. The manner in which it does so,
together with the conventional $\kappa$-mechanism, is discussed by 
Balmforth (1992a). It appears to have a significant destabilizing
influence on the pulsations (Balmforth \& Gough 1990a).

It was first reported by Gough (1980) that the dynamical effects arising 
from the turbulent momentum flux perturbations $\delta p_{\rm t}$ contribute 
significantly to the damping $\eta_{\rm t}$. Detailed analyses 
(Balmforth 1992a) reveal how damping is controlled largely by the phase 
difference between the turbulent pressure perturbation $\delta p_{\rm t}$ 
and the density perturbation $\delta \rho$.  Turbulent pressure fluctuations 
must not, therefore, 
be neglected in stability analyses of solar-type p modes.

\begfig 6.21 cm
\figure{2}
{Linear damping rates $\eta$ for the Sun as function of frequency.
The values chosen for the convection parameters are $\ac=1.8$ and 
$a^2=b^2=600$.}
\includegraphics{\figdir/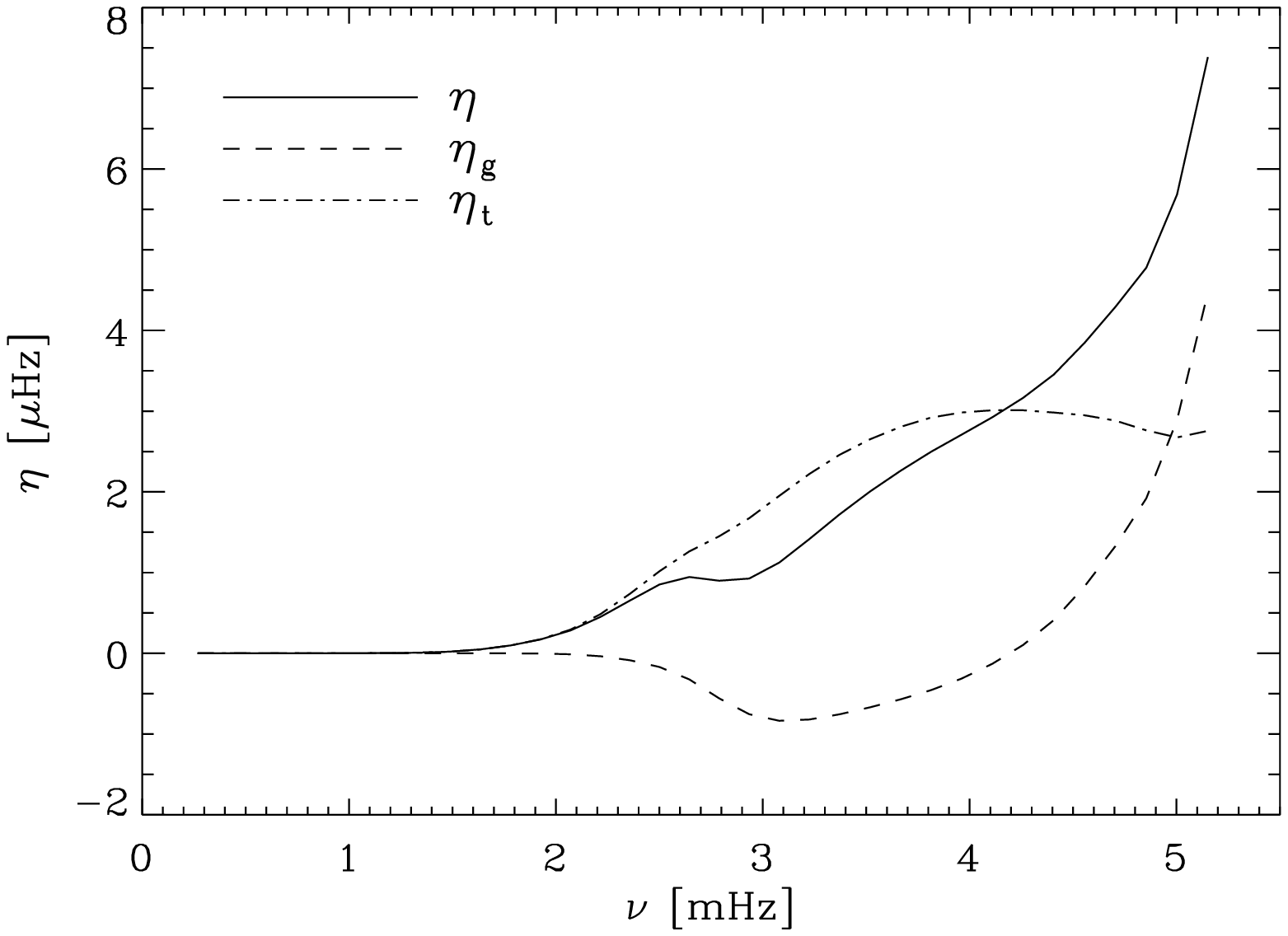}
\endfig

The results presented here were obtained from computations including 
the physics describing $\eta_{\rm rad}$, $\eta_{\rm conv}$ 
and $\eta_{\rm t}$. 
The nonadiabatic contributions $\eta_{\rm rad}$ and $\eta_{\rm conv}$
may be associated with the thermodynamics of the gas, and accordingly we
couple them into 
$\eta_{\rm g}=\eta_{\rm rad}+\eta_{\rm conv}$ (see also Fig.~1).

\vskip-3truemm
\titleb{Theoretical damping rates}
Damping rates are computed as the imaginary part $\omega_{\rm i}$ of 
the complex eigenfrequency $\omega=\omega_{\rm r}+{\rm i}\,\omega_{\rm i}$, 
obtained from solving the fully nonadiabatic pulsation equations. 
Balmforth (1992a) computed damping rates for the Sun, and reported that
he found all modes to be stable, with damping rates agreeing tolerably 
with observation for frequencies between $2$~mHz and $4$~mHz. Below and 
above this frequency range the theoretical damping rates are smaller than 
observations would suggest. Damping arising from incoherent scattering 
$\eta_{\rm scatt}$ (Goldreich \& Murray 1994, see Fig.~1), which may remove 
the discrepancy both at low and at high frequencies, is not modelled in 
our calculations.

Figure~2 displays the damping rates and their contributions arising from the
gas and turbulent pressure perturbations for a solar envelope model.
Damping is much augmented by the turbulent pressure perturbation 
$\delta p_{\rm t}$; it is only at the highest frequencies that the 
nonadiabatic contribution $\eta_{\rm g}$ to damping of solar p modes 
exceeds that from the turbulent pressure $\eta_{\rm t}$.

\begfig 6.21 cm
\figure{3}
{Damping rates for an evolving $1\,M_\odot$ star as function of frequency.
The results are displayed for models with 
ages=(0, 2.49, 3.96, 4.55, 6.19, 7.00, 8.03, 9.02, 9.72) Gy. The thick curve
indicates the results for the present Sun. Values $\ac=1.8, a^2=b^2=600$ 
for the convection parameters have been used.}
\includegraphics{\figdir/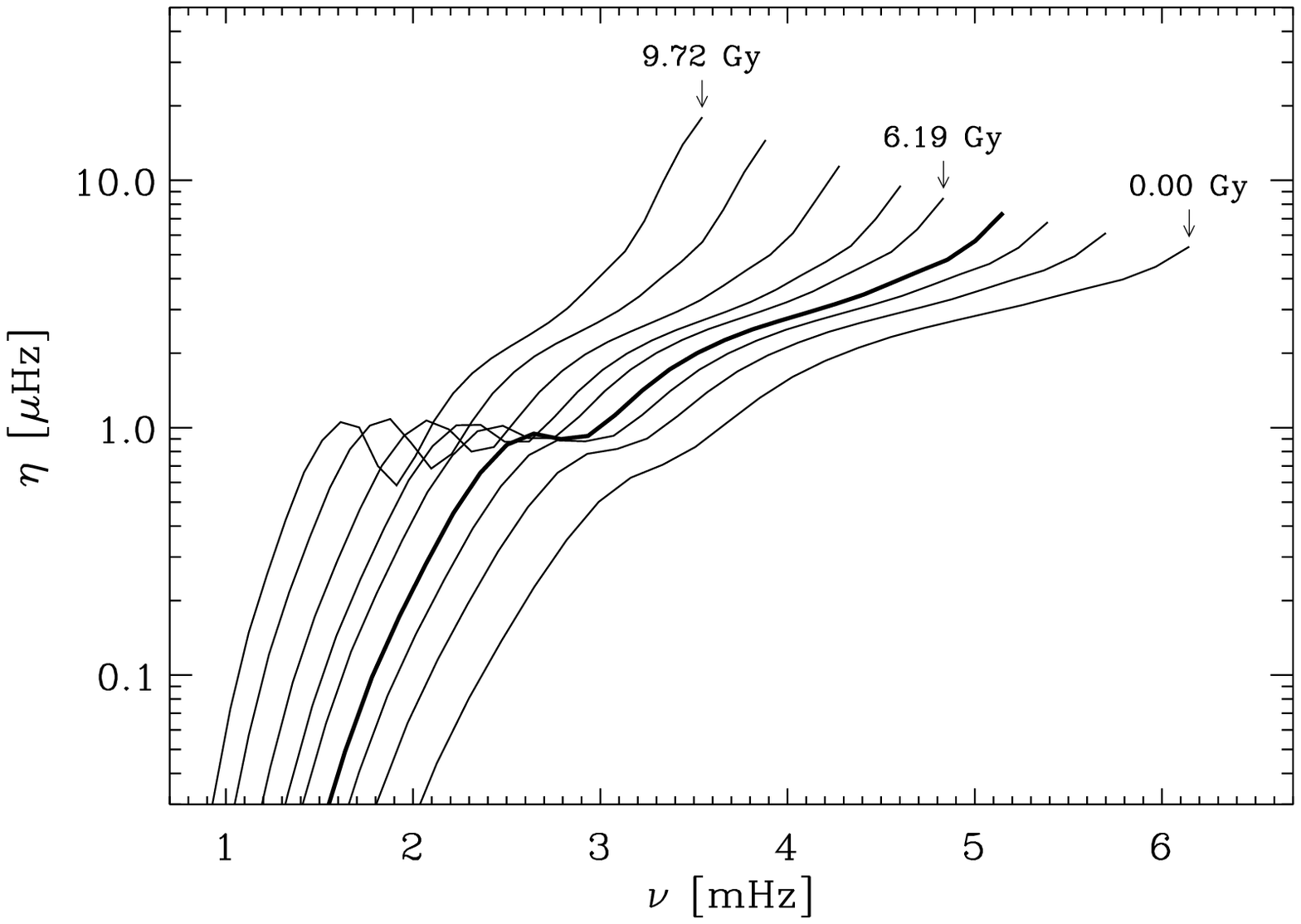}
\endfig

The total damping rate $\eta$ (solid curve), plotted as a function of cyclic
frequency $\nu=\omega_{\rm r}/2\pi$, is characteristically flat at frequencies
near 2.8~mHz (see Fig.~2). This feature is also observed in solar linewidth 
measurements
(e.g. Libbrecht 1988, Appourchaux\etal\ 1998, Chaplin\etal\ 1998).
At these frequencies the net damping is reduced particularly by
radiative processes in the upper superadiabatic boundary layer of
the convection zone, which are locally destabilizing.


\begfig 6.21 cm
\figure{4}
{Damping rates for an evolving $1.45\,M_\odot$ star as a function 
of frequency.  The results are depicted for models with 
ages=(0, 0.96, 1.38, 1.72, 2.00, 2.44) Gy. 
Values $\ac=2.0, a^2=900, b^2=2000$ for the convection parameters 
have been used.}
\includegraphics{\figdir/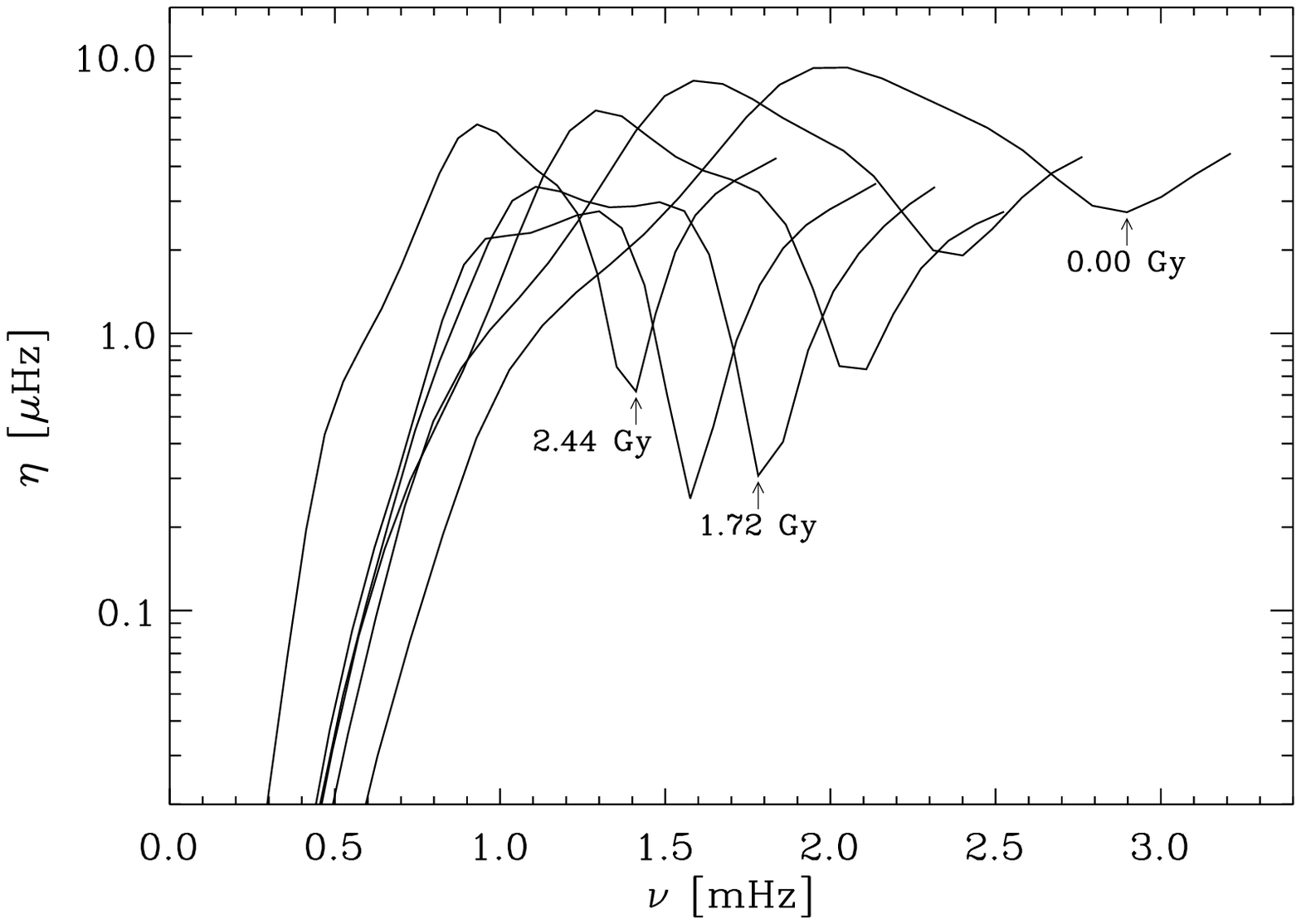}
\endfig

The damping rates for an evolving $1\,M_\odot$ star are depicted in Fig.~3. 
Damping rates generally increase with increasing age, particularly for low- 
and high-order modes. For modes of intermediate order the flattening of
the damping-rate curve becomes more pronounced as the star evolves, and 
turns into a locally concave function at about the solar age.
The maximum value of the superadiabatic temperature gradient of a 
$1\,M_\odot$ star increases by approximately $24\%$ along the main sequence, 
promoting the depression in the damping rates.

\begfig 6.21 cm
\figure{5}
{Damping rates for ZAMS models as functions of frequency.
The results are displayed for models with $M$=(0.95, 1.00, 1.05, 
1.10, 1.15, 1.20, 1.25, 1.30, 1.35, 1.40, 1.45, 2.00) $M_\odot$.  
For the convection parameters the values $\ac=2.0, a^2=900, b^2=2000$ 
have been used.}
\includegraphics{\figdir/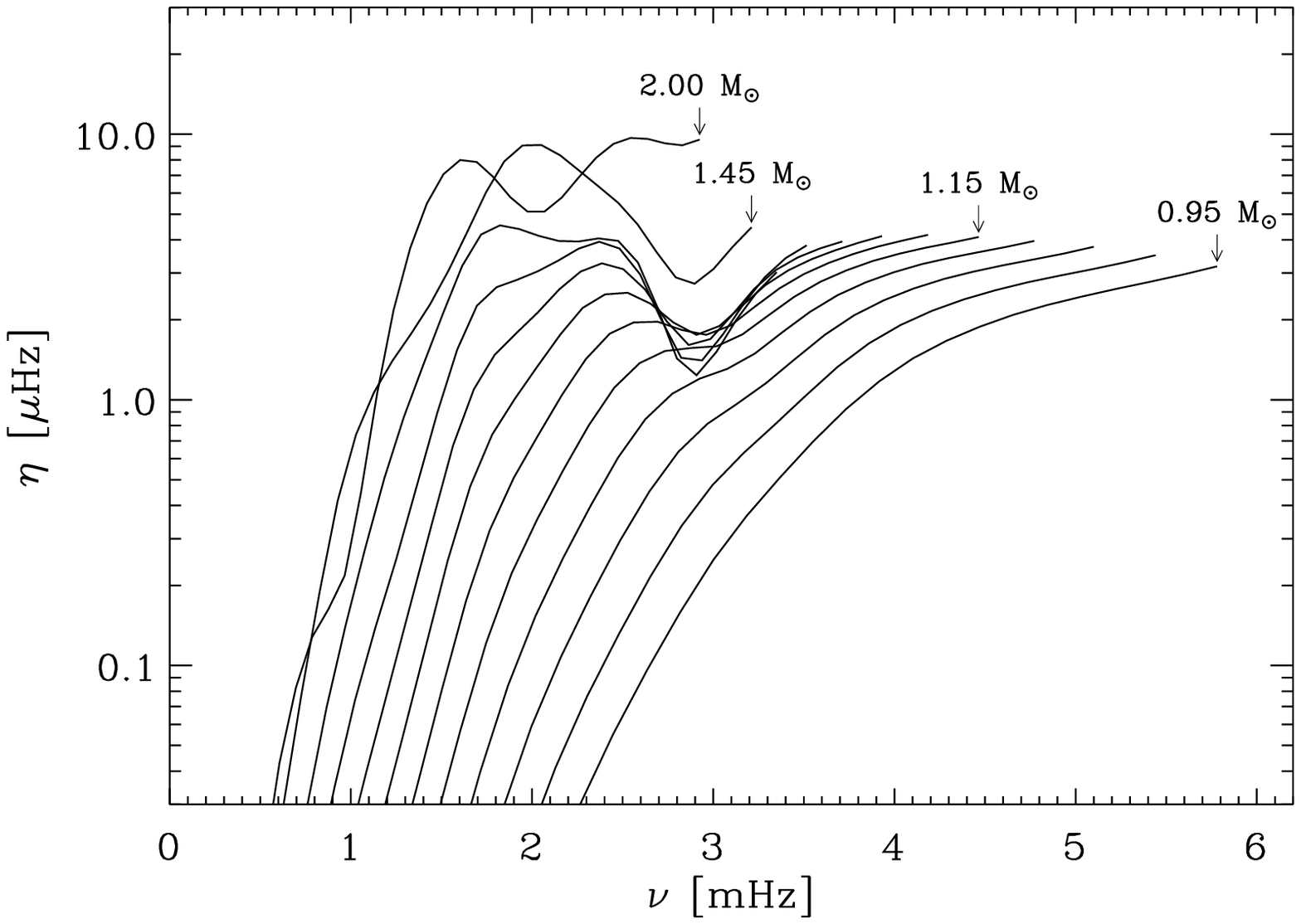}
\endfig

A similar behaviour of the damping rates is obtained for more massive stars,
as indicated for the evolving $1.45\,M_\odot$ star depicted in Fig.~4. 
For these more massive stars, larger values of the nonlocality 
parameters, $a^2=900$ and $b^2=2000$, were adopted. This selection 
ensured that all the radial modes were stable, in line with our working 
hypothesis that stochastic excitation underlies the appearance of solar-like 
oscillations (but see section 9).  The depression in the damping rates is more 
pronounced for these stars, even at the ZAMS. 
This trend may be seen even more obviously in 
Fig.~5, where damping rates are depicted for stars with increasing mass
along the ZAMS.

The functional dependence of $\eta$ on stellar parameters was determined
approximately by Goldreich \& Kumar (1991); they derived an 
order-of-magni\-tude estimate for the damping rates accounting roughly for 
the effects of radiative damping and convective dynamics. The radial 
modes were treated in the polytropic approximation to the outer layers
and convection was described by standard (unperturbed) local mixing-length
theory (B\"ohm-Vitense 1958). They obtained the expression
$$
  \eta\sim{L\over c_{\rm s}^2I_\omega}
          \Big({\omega_{\rm r}\over\omega_{\rm c}}\Big)^2
  \,,\eqno\autnum
$$
where $L$ is the luminosity and $c_{\rm s}$ denotes the adiabatic 
sound speed at the photosphere (which we define at the level where 
the temperature is equal to the effective temperature), $I_\omega$ is 
the mode inertia, and $\omega_{\rm c}$ is the acoustical cut-off 
frequency in an isothermal atmosphere (Lamb 1909),
$$
  \omega_{\rm c}\ist {c\over 2H_p}
  \,,\eqno\autnum
$$
where $c$ denotes the sound speed. 
The inertia is usually defined such that it represents the coefficient of 
proportionality between the energy in the mode and the square of the 
velocity amplitude of the associated disturbance in the surface layers 
of the star.  
The surface of the star, however, commonly lies in a region where the 
mode is evanescent, and in that case $I_\omega$ is more usefully regarded 
as a measure of evanescence, representing a property of the eigenfunction 
above the upper turning point (Gough 1995). For the case of a linear 
adiabatic mode of stellar 
oscillation, which can be represented by an undamped harmonic oscillator,
the mode inertia $I_\omega$ can be defined in terms of the total (kinetic + 
potential) energy $E$ and the mean-square value of its surface velocity 
$V_{\rm s}$, i.e., $E=I_\omega\,V_{\rm s}^2$, and consequently

$$
 I_\omega\ist{1\over\xi^2(R_\star)}\,
             \int_{m_{\rm b}}^{M_\star}|\xi(m,\omega_{\rm r})|^2\,\dd m
 \,.\eqno\autnum
$$
Here $m_{\rm b}$ denotes the mass interior to the bottom boundary of the 
envelope, and $R_\star$ and $M_\star$ represent respectively the radius and 
the mass of the star. In practice, we normalize the eigenfunction 
$\xi(m,\omega_{\rm r})$ such that
$$
  {\xi(R_\star)\over R_\star}\ist 1
  \,.\eqno\autnum
$$

In Fig.~6 we compare numerically computed damping rates with estimates from 
expression (2) with the right-hand side multiplied (arbitrarily) by 1/6. 
The factor 
1/6 can be obtained approximately by taking into account the adiabatic 
exponents ($\gamma_1$ and $\gamma_3$) in the derivation of expression (2), 
assuming a fully ionized gas. 
For the mode inertia $I_\omega$ (cf. Eq.\ 4) the calculations assumed 
adiabatically (dashed curves) and nonadiabatically (solid curves) computed 
displacement eigenfunctions $\xi$.

\begfig 6.01 cm
\figure{6}
{Theoretical damping rates as functions of frequency for the Sun, a
$1.25\,M_\odot$ and a $1.45\,M_\odot$ ZAMS star. The curves are
the right-hand side of expression (2) multiplied by the factor $1/6$, 
assuming adiabatically (dashed curves) and 
nonadiabatically (solid curves) computed mode inertia $I_\omega$. 
The symbols show the damping rates obtained directly from the 
corresponding pulsation calculations, in which were solved the fully 
nonadiabatic linearized equations using the nonlocal, time-dependent 
mixing-length model with the convection parameters of Figs~4 and 5.}
\includegraphics{\figdir/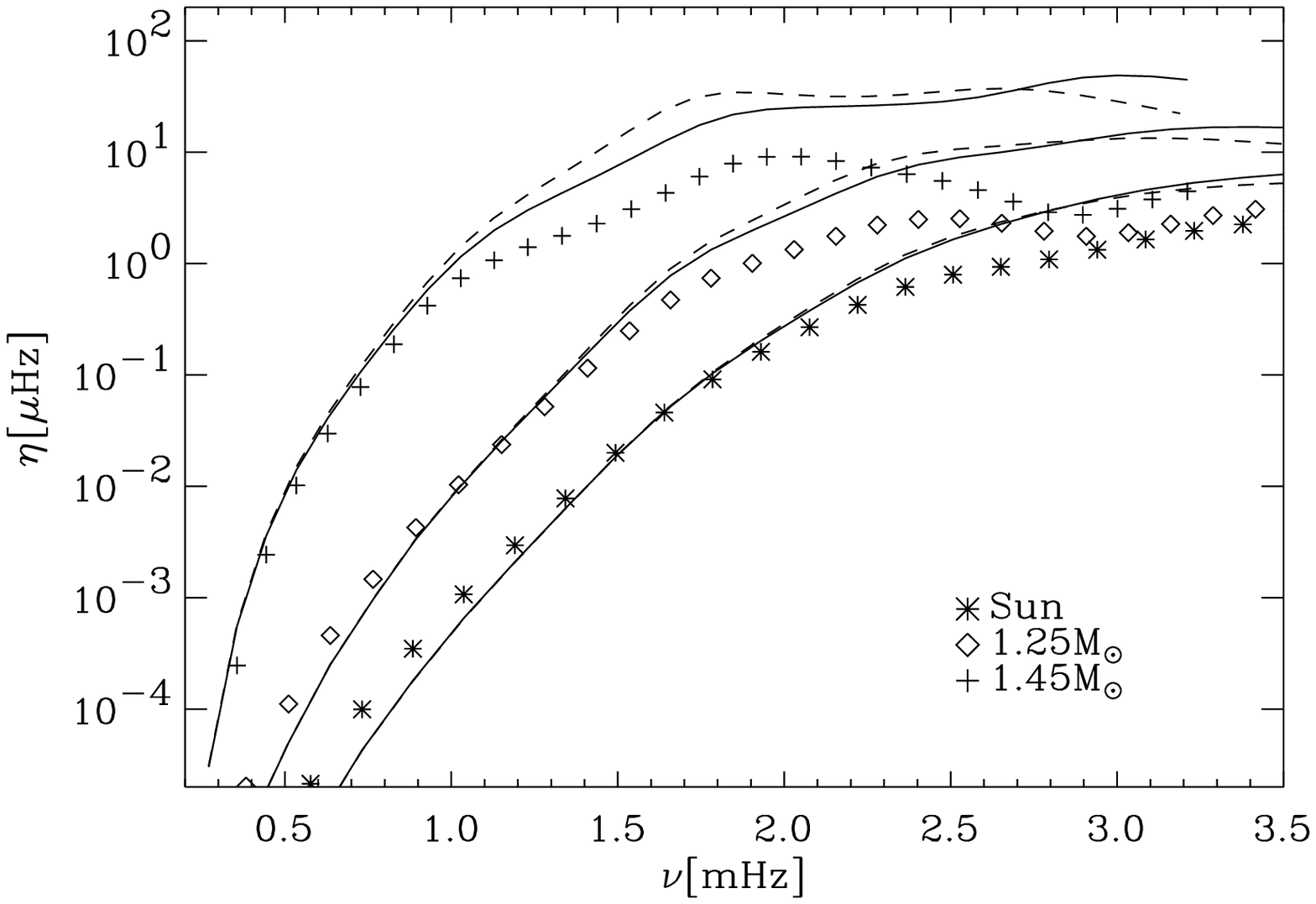}
\endfig
 
For frequencies below about half the isothermal acoustical 
cut-off frequency $\omega_{\rm c}$, the results 
suggest a fair agreement between analytical and modelled damping rates.
An interesting feature is the bend in the analytical solution obtained with
both the nonadiabatic and adiabatic eigenfunctions for the $1.45\,M_\odot$ 
star near the frequency $\nu \simeq 1.85$~mHz. This property is obviously 
related to the shape of the eigenfunctions in the boundary layers of the
convection zone, because in a polytrope the mode inertia $I_{\rm\omega}$ 
is a smooth function of height (Gough 1995). The characteristic flattening
of the damping rates (e.g., near 2.8~mHz for the Sun), however, is not 
seen in the estimates from expression (2).

\vskip -3truemm
\titlea{Amplitude ratios}

\putattop\begfig 6.20 cm
\figure{7}
{Theoretical amplitude ratios for a solar model 
compared with observations by Schrijver\etal\ (1991). 
Computed results are depicted for velocity amplitudes obtained at 
different heights above the photosphere ($h=0$\,km at $T=T_{\rm eff}$) 
assuming the convection parameters used for Figs~2 and 3.
The thick, solid curve indicates a running-mean average of the data.}
\includegraphics{\figdir/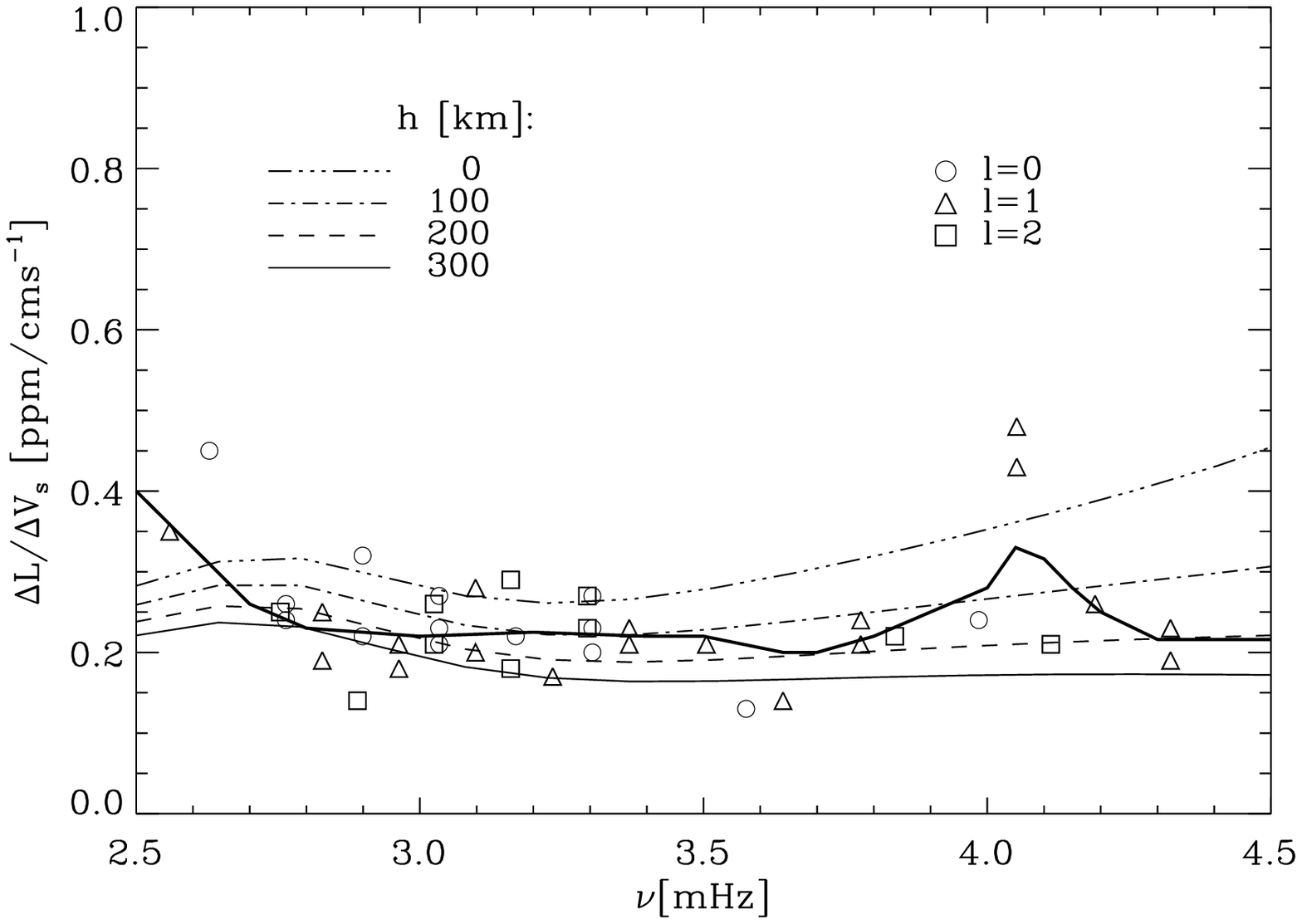}
\endfig

A useful test of the pulsation theory, independent of an excitation model,
is provided by comparing estimated intensity-velocity amplitude ratios with 
observations.
For the Sun, accurate irradiance measurements exist from the IPHIR 
instrument of the PHOBOS 2 spacecraft with contemporaneous low-degree 
velocity observations in the potassium line 
from the Birmingham instrument at Tenerife (Schrijver\etal\ 1991).
This allows us to compare observed solar amplitude ratios 
with our estimated ratios as function of frequency. The comparison is
displayed in Fig.~7, where the model results are depicted for velocity 
amplitudes computed at different atmospheric levels. For moderate and 
high eigenfrequencies the amplitude of the displacement $\xi$ increases 
quite steeply in the evanescent outer region of the atmosphere, where 
the density declines very rapidly. The computed velocity amplitude, and 
hence the ratio, varies by about 15\,\% between the photosphere and the 
temperature minimum.  Thus attention has to be paid
to which atmospheric level the velocity amplitudes are computed, i.e.,
at which level the displacement eigenfunctions are normalized.
Observations are performed in selected Fraunhofer lines, e.g. in the 
neutral potassium line ($769.9$~nm) as used in the BiSON observation
(Elsworth\etal\ 1993), which is formed at a height of $h\simeq 200$\,km 
above the point where the temperature is equal to the effective 
temperature (assuming the $T$-$\tau$ relation derived from the model C 
atmosphere of Vernazza\etal\ 1981). 
The luminosity amplitudes have been computed at the
outermost gridpoint and a correction factor has been applied to account for 
the conversion to the measured irradiance wavelength of $\lambda=500\,$nm 
using the approximation of Kjeldsen \& Bedding (1995). Observations with a 
coherence greater than $0.7$ are represented by different symbols denoting 
measurements of different degree $l$. The thick solid curve represents a 
running-mean average, with a width of 300\,$\mu$Hz, of the observational 
data. The theoretical ratios for $h=200$ km (dashed curve) show reasonable 
agreement with the observations.

\vskip -3mm
\titlea{Acoustical noise generation rate}
Acoustical radiation by turbulent multipole sources in the context of stellar
aerodynamics has been  considered by Unno \& Kato (1962), Moore \& 
Spiegel (1964), Unno (1964), Stein (1967),  Goldreich \& Keeley (1977b), 
Osaki (1990), Balmforth (1992b), Goldreich, Murray \& Kumar (1994) 
and Musielak\etal\ (1994).

In a pulsating atmosphere the full pulsation-convection
equations must be derived from the fluid-dynamical equations in which 
the fluid velocity includes both turbulence and pulsation. Balmforth (1992b) 
reviewed the theory of acoustical excitation in a pulsating atmosphere, and, 
following Goldreich \& Keeley (1977b), he derived the following expression 
for the rate of energy injected into a mode with frequency $\omega_{\rm r}$ 
by quadrupole emission through the fluctuating Reynolds stresses:
$$
\eqalign{
P_{\rm Q}\ist &{\pi^{1/2}\over 8I_\omega\xi^2(R_\star)}\cr
        &\times
        \int\limits_{M_\ast}
        \left({\partial\xi(m,\omega_{\rm r})\over\partial r}\right)^2
        \rho \ell_0^3u_0^4\tau_0{\cal S}(m,\omega_{\rm r})\,\dd m\,,
        }
        \eqno\autnum
$$
where $\ell_0$,$u_0$,$\tau_0$ are respectively the length, velocity and
correlation time scales of the most energetic eddies, determined by the 
mixing-length model. The function ${\cal S}(m,\omega_{\rm r})$ accounts
for the turbulent spectrum, which approximately describes contributions
from eddies with different sizes to the noise generation rate $P_{\rm Q}$,
and which we implemented as did Balmforth (1992b):
$$
{\cal S}(m,\omega_{\rm r})=\int_0^\infty{u_\kappa^3\over\kappa^5}
        \exp[-\omega_{\rm r}^2\tau_0^2/(2\kappa u_\kappa)^2]\,\dd\kappa
\,,\eqno\autnum
$$
where $\kappa=k/k_0$, $u_\kappa=u(k)/u_0$, $k$ is the wavenumber of 
an eddy with
velocity $u(k)$, and $k_0$ is the wavenumber at the peak of the spectrum.
For the computation of $u(k)$, a turbulent spectrum according to 
Spiegel (1962) has been chosen.

The emission of acoustical radiation by turbulent multipole sources
depends critically on the convective velocity $u$. In homogeneous isotropic
and non-decaying turbulence, acoustic emission by the fluctuating 
Reynolds stresses (quadrupole emission) scales with the fifth power 
of the turbulent Mach number $M_{\rm t}=u/c$ (Lighthill-Proudman formula). 
Inhomogeneity and anisotropy effects in the overturning layers of stars
give rise to monopole and dipole emission manifested in the fluctuation of
the entropy (e.g. Goldreich \& Kumar 1990). 
Stein \& Nordlund (1991) and Goldreich, Murray \& Kumar (1994)
suggest that the monopole and dipole source may be as important
as quadrupole radiation. 
However, previous work has demonstrated
that the prescription is capable of roughly reproducing solar
measurements, and so, partly for want of a serious theory, we stick
with the expressions (6) and (7) here.


\vskip -2truemm
\titlea{Amplitudes}
\titleb{Velocity amplitudes}

\putattop\begfig 6.21 cm
\figure{8}
{Velocity amplitudes for the Sun as a function of frequency.
The computed values (continuous curve) are depicted at the photospheric 
level $h=200$\,km. The turbulent spectrum ${\cal S}(m,\omega_r)$, given by
Eq.\ (7), has been multiplied by the factor 6.55 to fit the 
maximum value of the velocity data (filled circles) from the BiSON 
observations (Chaplin\etal\ 1998). The data are from a 32-month almost
continuous sequence collected between May 1994 and January 1997, i.e., 
at or near the solar-cycle 22/23 activity minimum. 
The computations assumed the convection parameters $\ac=1.8, a^2=b^2=300$.}
\includegraphics{\figdir/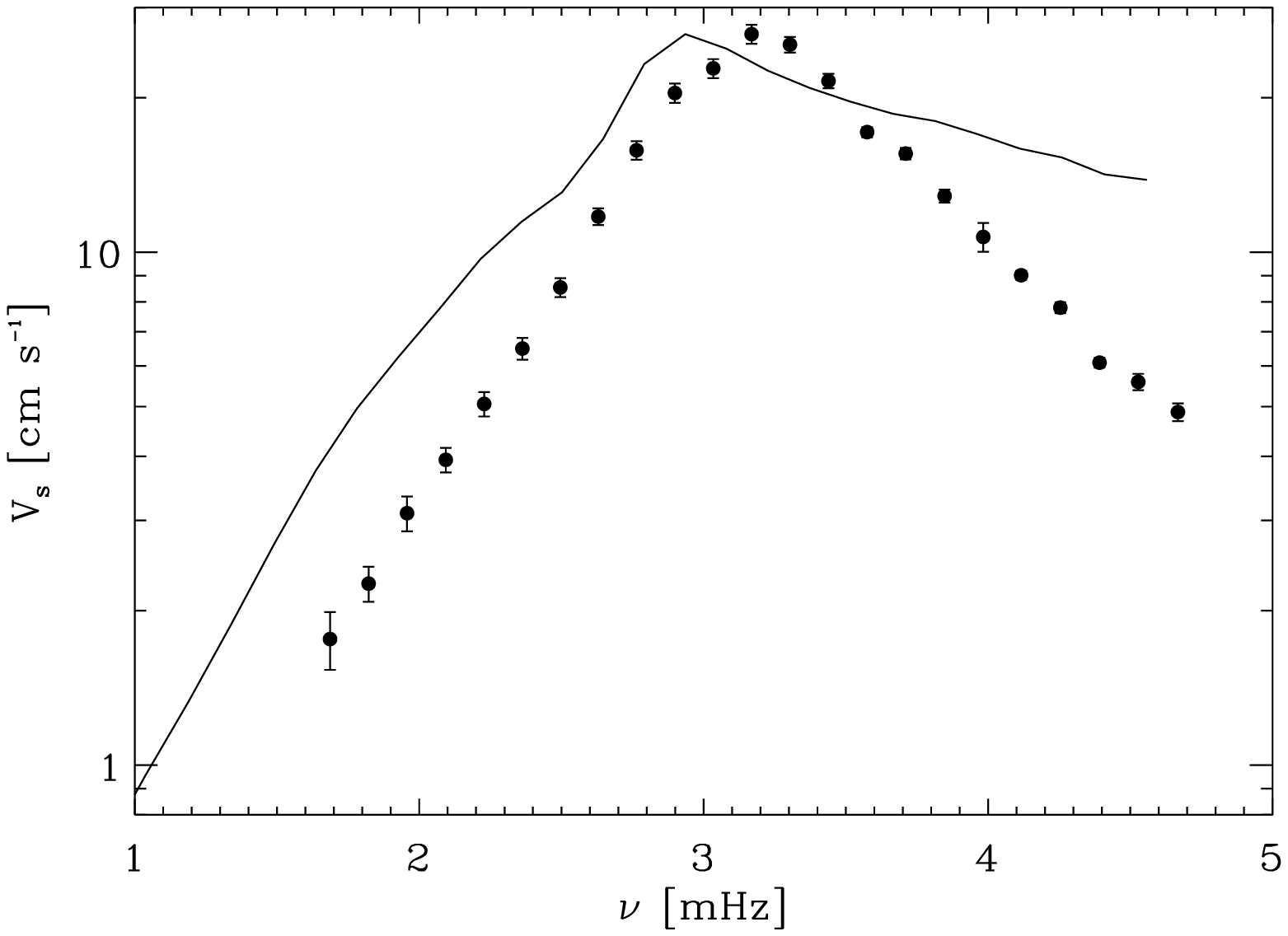}   
\endfig

\begfig 6.21 cm
\figure{9}
{Velocity amplitudes for an evolving $1\,M_\odot$ star 
as function of frequency, depicted at the photospheric level $h=200$\,km. 
The results are displayed for the model masses and convection parameters
of Fig.~3. The thick curve indicates the results for the Sun.}
\includegraphics{\figdir/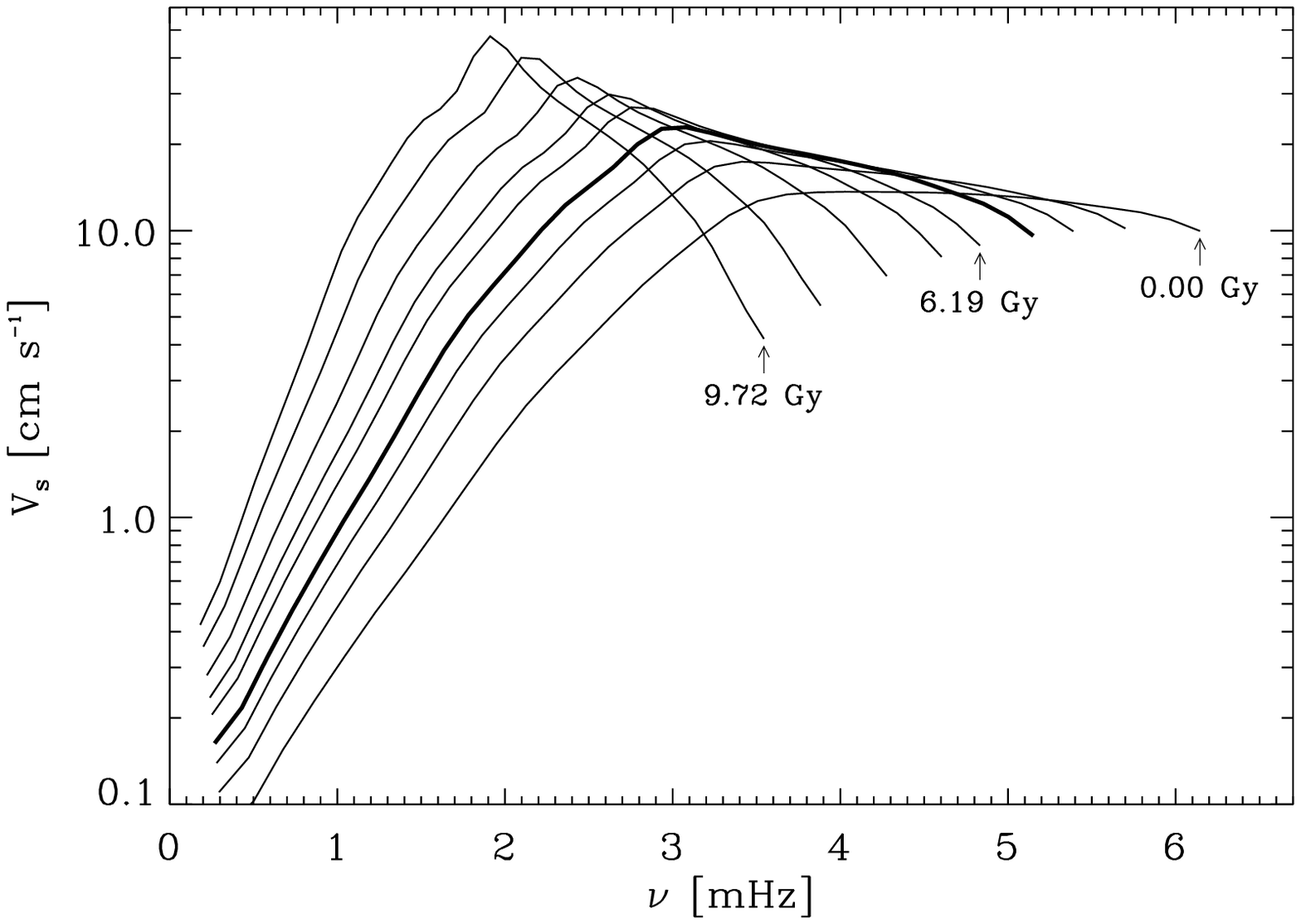}
\endfig

With the computations of the damping rate, $\eta$,
and noise generation rate, $P_{\rm Q}$, the root-mean-square velocity
at a particular level in the atmosphere may be written as
$$
V_{\rm s}
   \ist\,\sqrt{P_{\rm Q}\over 2\,\eta\,I_\omega}
  \,.\eqno\autnum
$$

\begfig 6.21 cm
\putattop\figure{10}
{Velocity amplitudes for an evolving $1.45\,M_\odot$ star as function 
of frequency, computed at a height $h=200$\,km. The dashed curve displays
the result for the 2.44\,Gy model applying a median filter on the 
amplitudes with a width corresponding to nine radial modes.
The amplitudes are portrayed for the model ages and convection parameters 
of Fig.~4.}
\includegraphics{\figdir/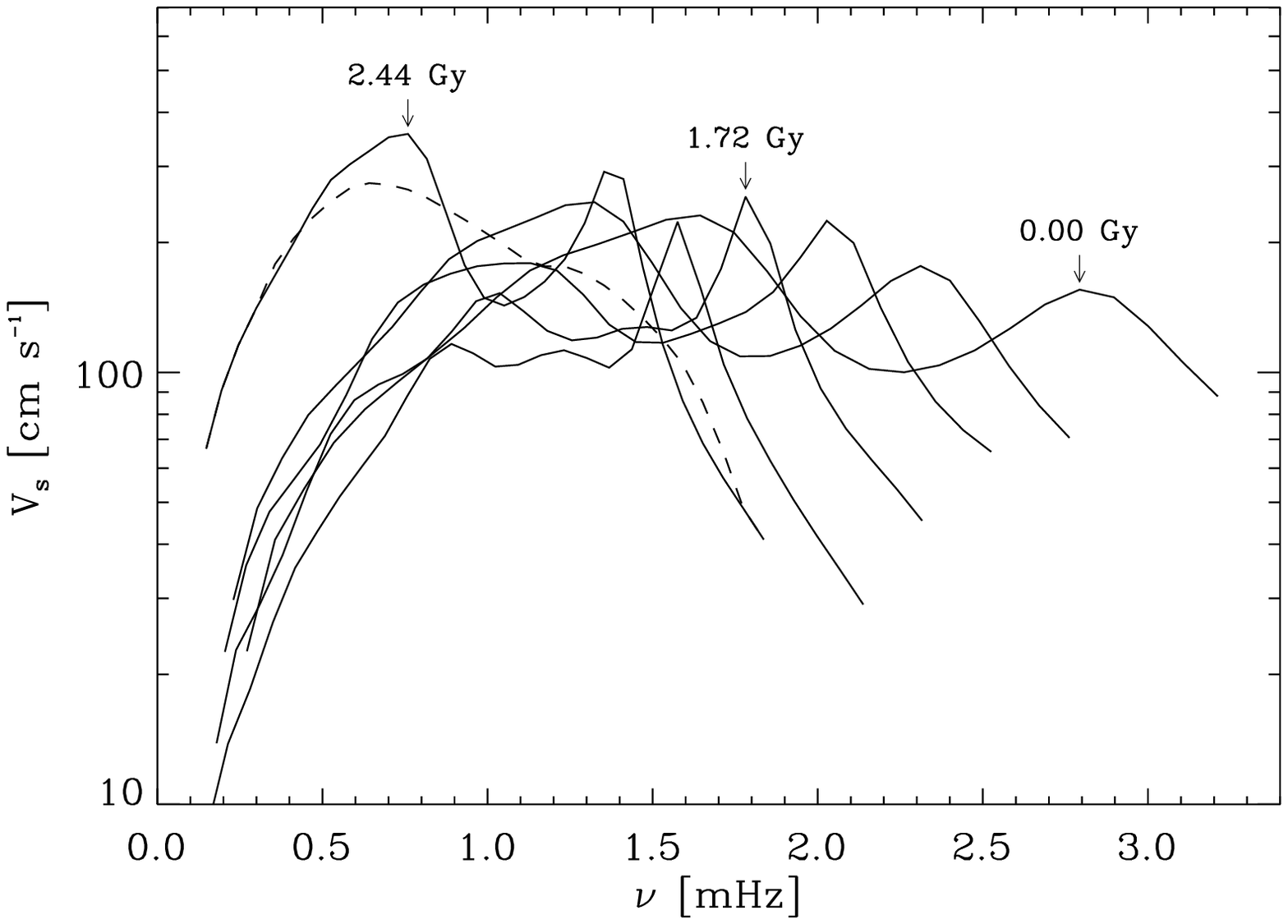}
\endfig

The form of the turbulent spectrum ${\cal S}(m,\omega_{\rm r})$, given 
by Eq.\ (7), has a substantial effect on the predicted mode amplitudes. In 
this paper we 
multiplied the rhs of Eq.\ (7) by the factor 6.55 for all amplitude 
predictions. This empirical correction, which can be attributed perhaps to
uncertainties in our expressions for the quadrupole emission, leads to
theoretical solar velocity amplitudes that have the same maximum value of
26.6\,cm\,s$^{-1}$ as that observed by the BiSON group (Chaplin\etal\ 1998). 
The results of the scaled theoretical mean amplitude values for a solar model 
are displayed in Fig.~8 together with the BiSON data. 

\titlec{Main-sequence stars}
The mean velocity as a function of frequency, computed at a height 
$h=200$\,km above the photosphere of an evolving $1\,M_\odot$ star, 
is displayed in Fig.~9. 
The oscillation amplitudes become larger with age for low and intermediate
frequencies, exhibiting a maximum value of 
$V_{\rm s}\ \simeq $\,45\,cm\,s$^{-1}$ 
at the end of the hydrogen core-burning phase.
The increase comes about because the ratio $P_{\rm Q}/I_\omega$ increases 
with age at the frequency of maximum mode energy, whereas the damping rates
decrease with age at this frequency (see Fig.~3).

In Fig.~10 the amplitudes are depicted for an evolving $1.45\,M_\odot$ star, 
also computed at the height $h=200$\,km. For models before the characteristic 
`hook' (i.e., at ages $\la$ 2.36\,Gy)
in the evolutionary track (see Fig.~13) the amplitudes 
increase only moderately with age, mainly because of the increasing
mode inertia at the frequency of maximum mode energy and the consequent
decrease of the ratio $P_{\rm Q}/I_\omega$. For models older than
$\sim$\,2.36\,Gy, the luminosity increases fairly rapidly, leading to
a steep increase in the turbulent Mach number and hence in the noise 
generation rate $P_{\rm Q}$, and consequently mode amplitudes.
Near the end of the hydrogen core-burning phase the theory predicts maximum 
values of $\sim$\,330\,cm\,s$^{-1}$ for the velocity amplitudes.
In general these maximum values coincide with the sharp depression in the 
damping rates (see Fig.~4).

\begfig 5.96 cm
\figure{11}
{Luminosity amplitudes for ZAMS model as function of frequency, displayed
at the outermost meshpoint of the models. The computations assumed the 
convection parameters and model masses of Fig.~5.}
\includegraphics{\figdir/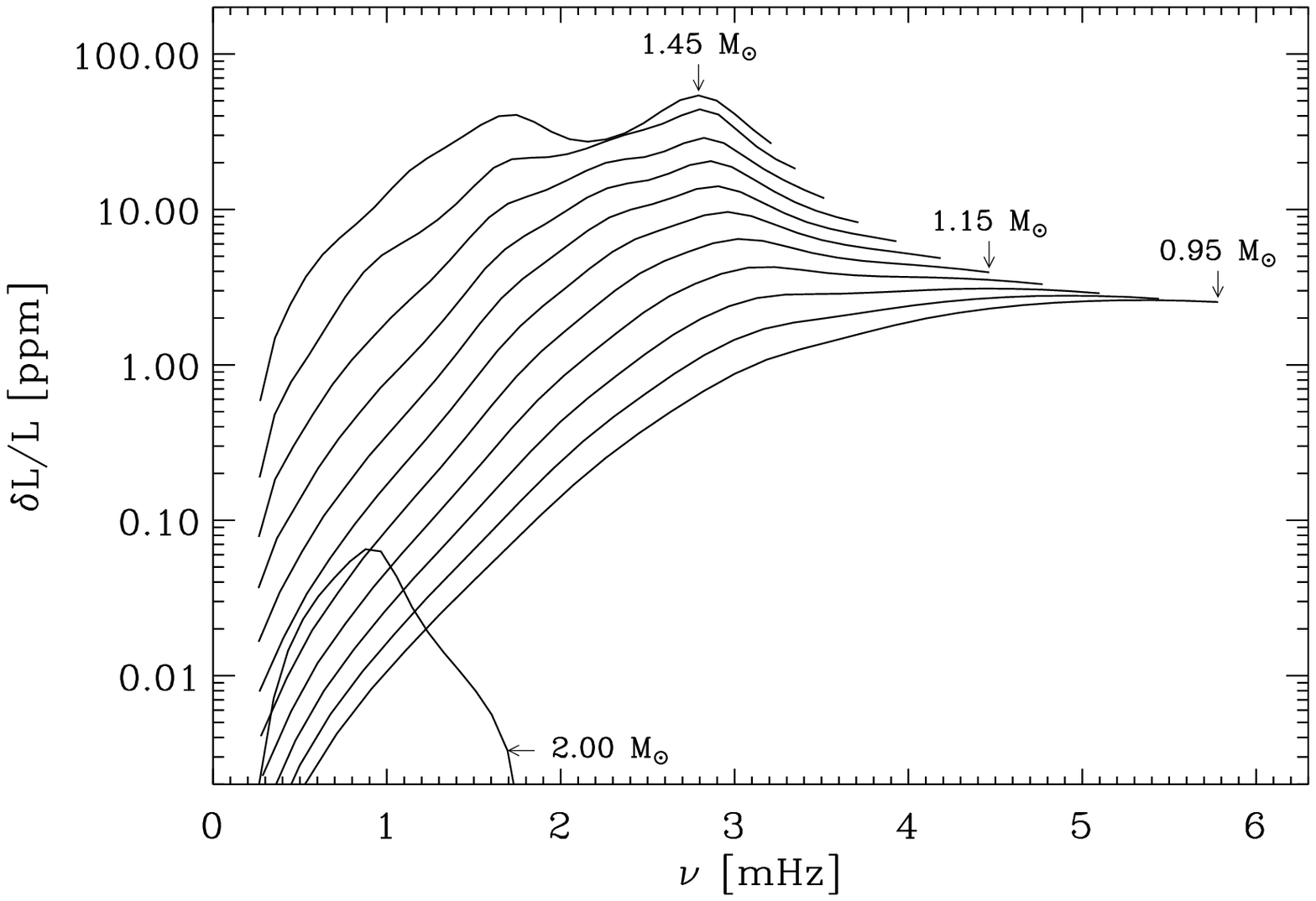}      
\endfig

\titleb{Luminosity amplitudes}

For radial modes,
the imaginary part of the nonadiabatic displacement eigenfunctions
is very small relative to the real part. The differences in the 
velocity amplitudes when using the adiabatic instead of the nonadiabatic 
displacement eigenfunctions are negligible relative to the uncertainties 
inherent in modelling the theory of stochastic excitation.
For the estimation of the luminosity amplitudes, however,
nonadiabatic eigenfunctions of the relative luminosity fluctuations,
$\delta L/L_0$, have to be taken into account. 
The relative luminosity amplitudes are related linearly to the velocity
amplitudes.

\begfig 5.78 cm
\figure{12}
{Maximum values of turbulent Mach number $M_{\rm t}$, turbulent pressure 
fraction $p_{\rm t}/p$ and convective growth rate $\sigma=2w/\ell$ as 
function of model mass along the ZAMS. The computations assumed the 
\hbox{convection parameters of Figs~4 and 5.}}
\includegraphics{\figdir/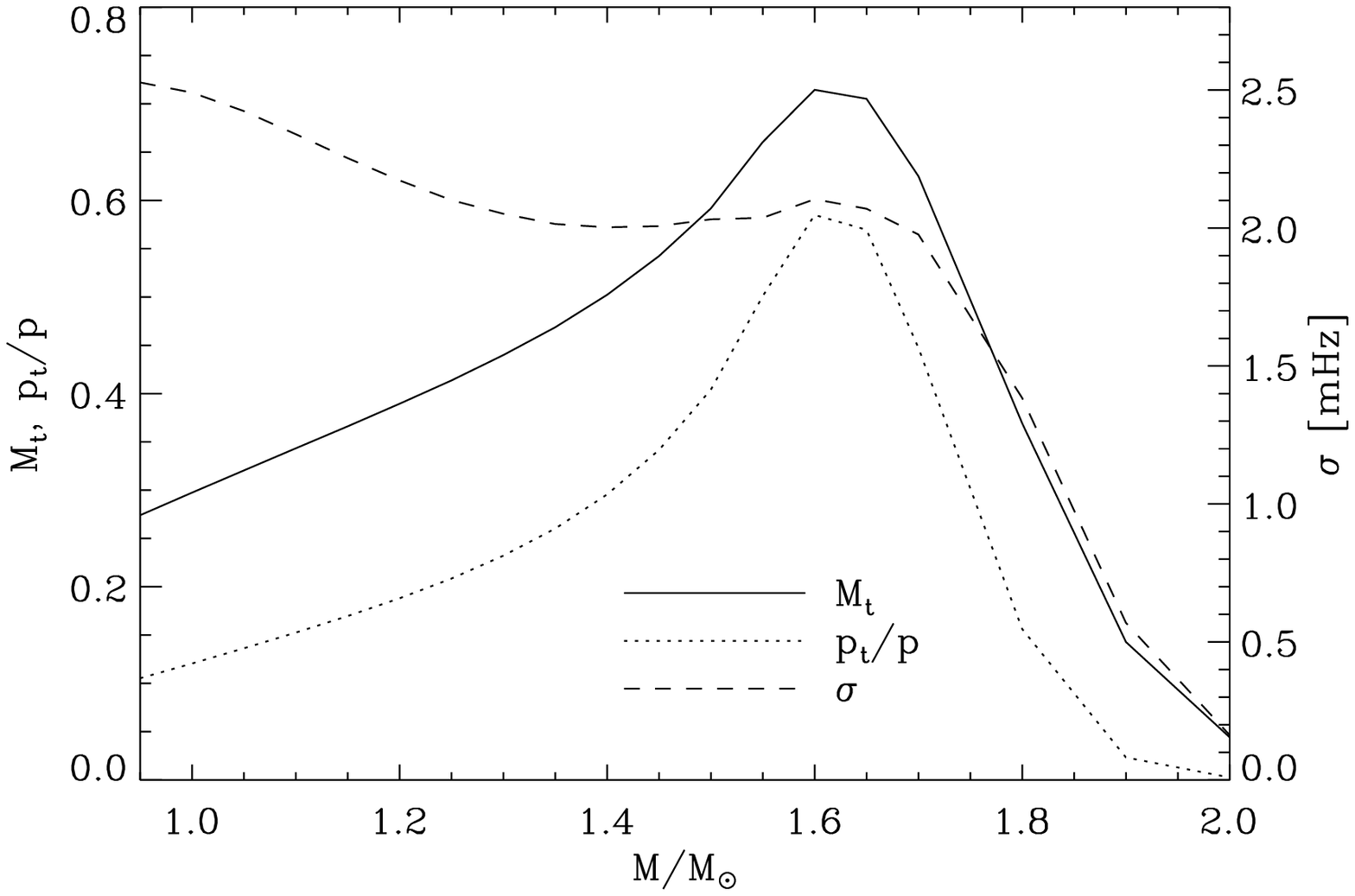}
\endfig

\titlec{Zero-age main-sequence stars}
The luminosity amplitudes of moderate-mass stars along the ZAMS
are depicted in Fig.~11.
The amplitudes increase monotonically with $M$ for stars 
with $M\le1.45\,M_\odot$, up to a maximum value of $\sim 50$\,ppm.
For models with $M\,\ga1.6\,M_\odot$, amplitudes of stochastically 
excited modes decrease with $M$; for a $2\,M_\odot$ ZAMS star the maximum 
amplitude is $\sim0.06$\,ppm. The dependence of the amplitude variations 
upon mass, or upon luminosity, may be explained principally by the strong 
dependence of the acoustic noise generation rate on the turbulent Mach 
number $M_{\rm t}$. 
The dependence of the maximum values of the turbulent Mach number 
$M_{\rm t}$ and the ratio of turbulent pressure to total pressure 
$p_{\rm t}/p$ upon model mass along the ZAMS is illustrated in Fig.~12. 
The computations predict the 
largest turbulent Mach numbers for models with a mass of $\sim 1.6\,M_\odot$.
The $2\,M_\odot$ ZAMS star exhibits two very thin convection zones
in the outer part of the envelope, and the theory predicts a maximum 
turbulent Mach number $M_{\rm t}<0.1$. 
Furthermore, the opacity and consequently the convective heat flux 
decrease with $M$ for $M\,\ga1.6\,M_\odot$.
Also indicated in Fig.~12 is the maximum value of the convective 
growth rate $\sigma=2w/\ell$, scaled in units of cyclic frequency $\nu$. 
The ratio $\sigma/\nu$ influences 
the shape of the eigenfunctions in such a way as to cause a local depression 
in the damping rates $\eta$ considered as functions of $\nu$ (cf. Gough 1997). 
The maximum value of $\sigma$ is roughly equal to the frequency of the local
minimum of $\eta$.

\begfigwid 12.91 cm
\figure{13}
{Unstable modes and mean velocity amplitudes of stochastically excited
oscillations. Amplitudes, evaluated at a height $h$=200\,km are depicted 
as contours (solid curves) labelled at the amplitude
values  19, 27, 35, 45, 60, 80, 110, 150, 250, 350 cm\,s$^{-1}$. 
The dotted curves are evolutionary tracks. 
The Sun, indicated by its symbol $\odot$, exhibits a mean (rms) velocity of
$20.0$~cm\,s$^{-1}$. Calculations have been carried out till the end of 
hydrogen core-burning, giving the low-temperature extremities of the 
contours. The location of the instability strips for the $n=1$ and $n=2$ 
radial modes are indicated by solid and dashed straight lines, respectively.}
\includegraphics{\figdir/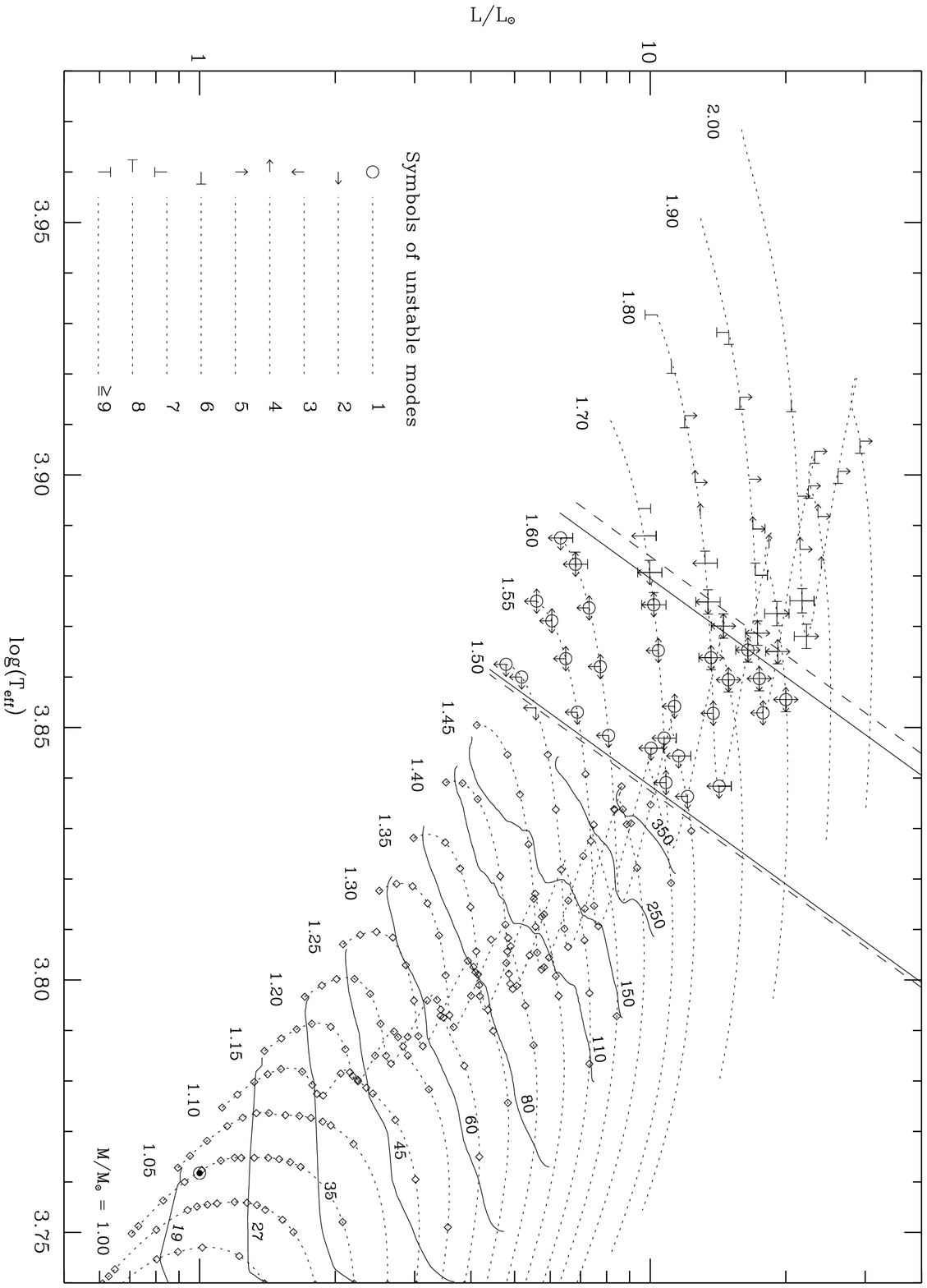}
\endfig

\titleb{Dependence on stellar parameters}
For those models on the main sequence for which all the modes are
predicted to be stable, the computed maximum velocity amplitudes
of stochastically excited modes (evaluated at the height $h=200$\,km) are 
shown as contours on the HR diagram in Fig.~13.
The $191$ models (indicated by the diamond symbols) were generated by 
specifying the mass, luminosity and effective
temperature provided from full evolution sequences, as obtained by
Christensen-Dalsgaard (1993). The same convective parameters as those in
Figs~4 and 5 were adopted. For more massive stars the maximum amplitudes
exhibit peaks in their frequency spectrum due to the sharp dip
in their damping rates (see Figs~4 and 10). We moderated these
peaks by applying a median filter to the amplitudes of all models with a 
width in frequency corresponding to nine radial modes (as illustrated
in Fig.~10 by the dashed curve for a 1.45\,$M_\odot$ 
star with an age of $2.44$\,Gy).

\begfig 12.27 cm
\putattop\figure{14}
{Luminosity amplitudes (top) and amplitude ratios (bottom) as function 
of effective temperature and model luminosity. The amplitude ratios are 
displayed for velocities at two different atmospheric levels:
the thin curves denote 
the results at $h=0$\,km and the thick curves at the height $h=200$\,km. 
The luminosity amplitudes are computed at the outermost meshpoint of the 
models.  The computations assumed model parameters as given in Fig.~4. 
The value for the Sun (3.4\,ppm) is indicated by its symbol.}
\includegraphics{\figdir/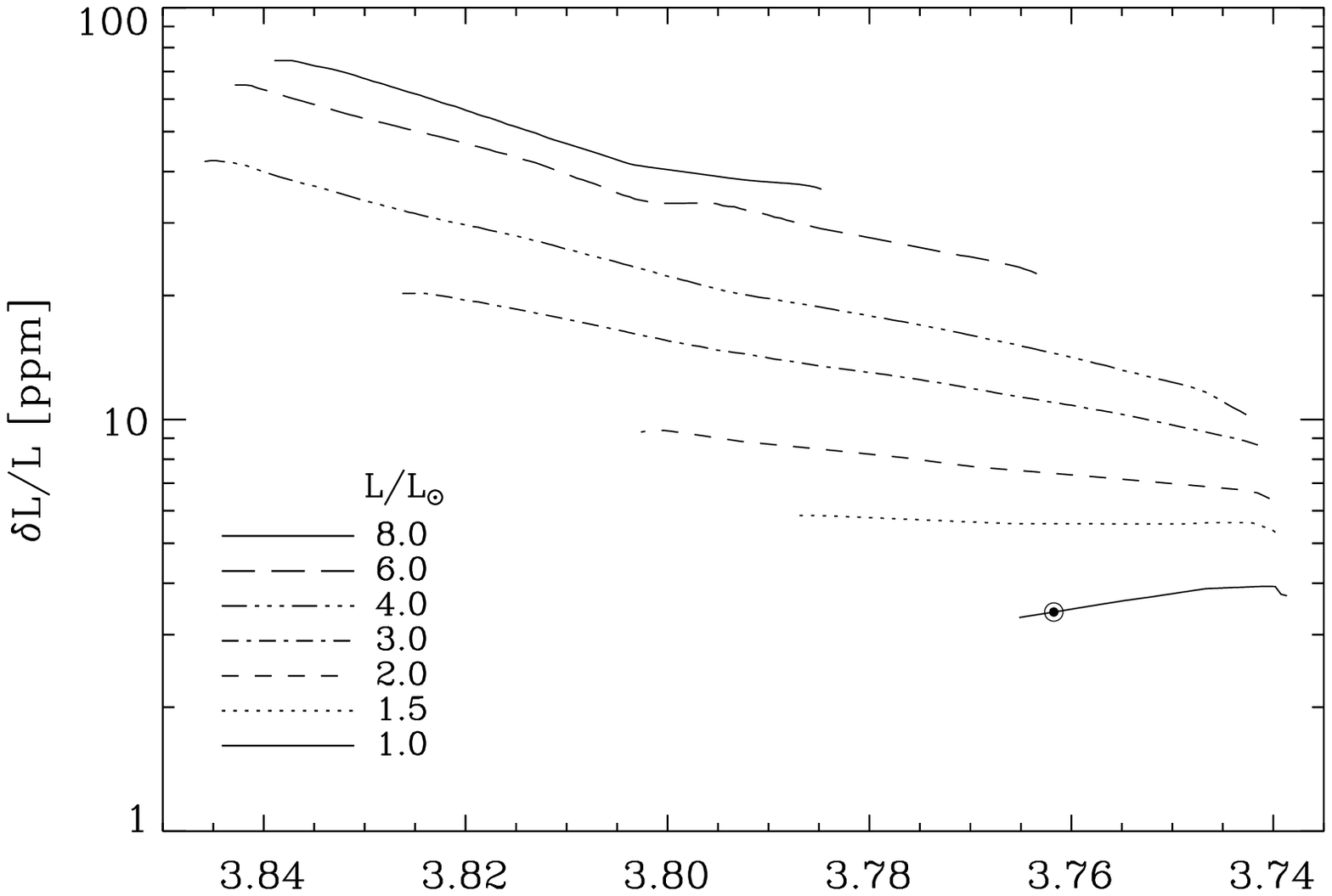}     
\includegraphics{\figdir/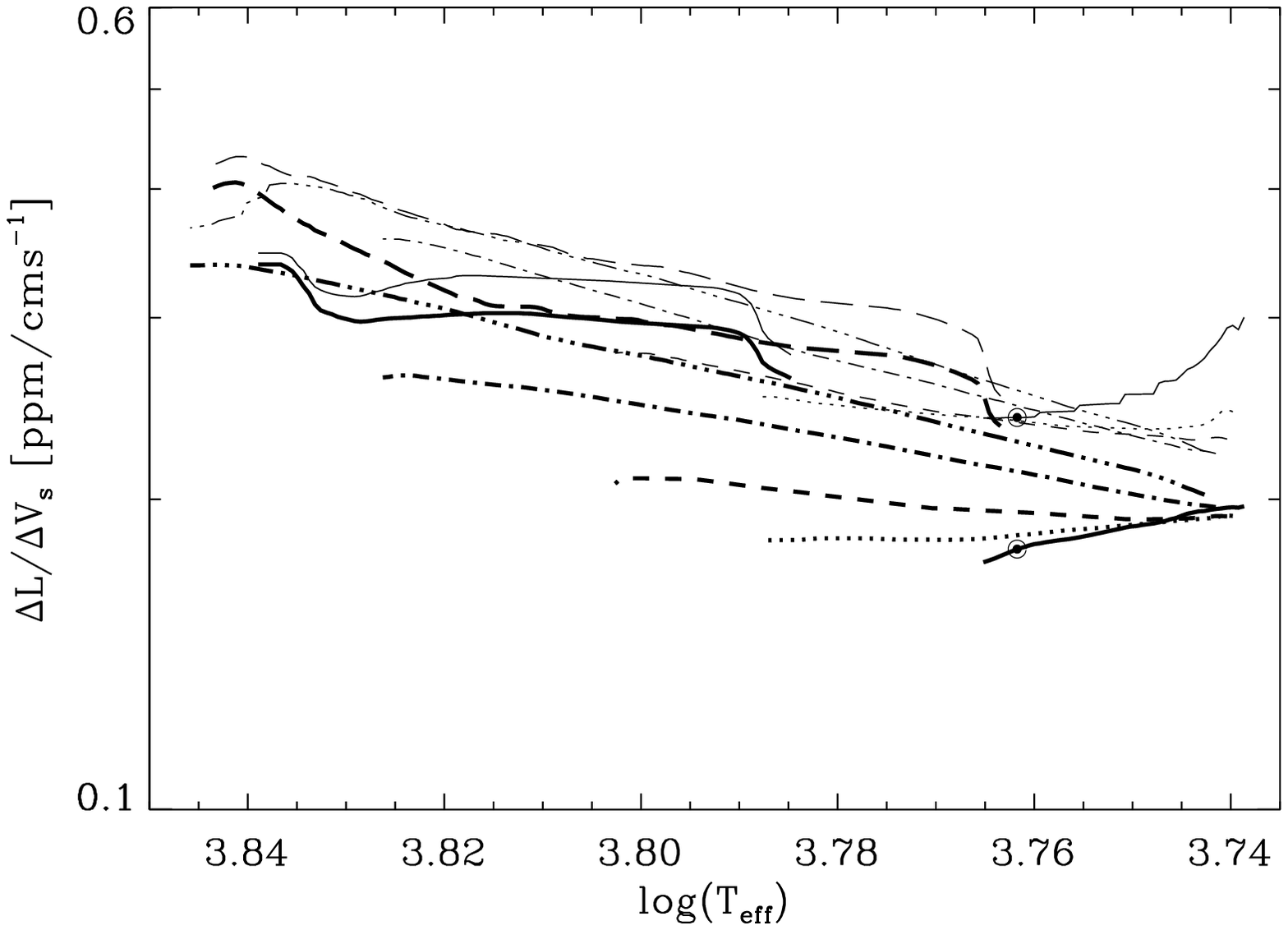}    
\endfig

\begfig 6.35 cm
\figure{15}
{Velocity amplitudes as function of light-to-mass ratio for
stochastically excited oscillations in 191 models (filled circles) 
calculated at a height $h$=200\,km above the photosphere with the same 
convection parameters as for Fig.~4. The dashed curve indicates Kjeldsen 
\& Bedding's scaling law (9) with $s=1$. The amplitudes are displayed 
relative to the value found in the Sun.}
\includegraphics{\figdir/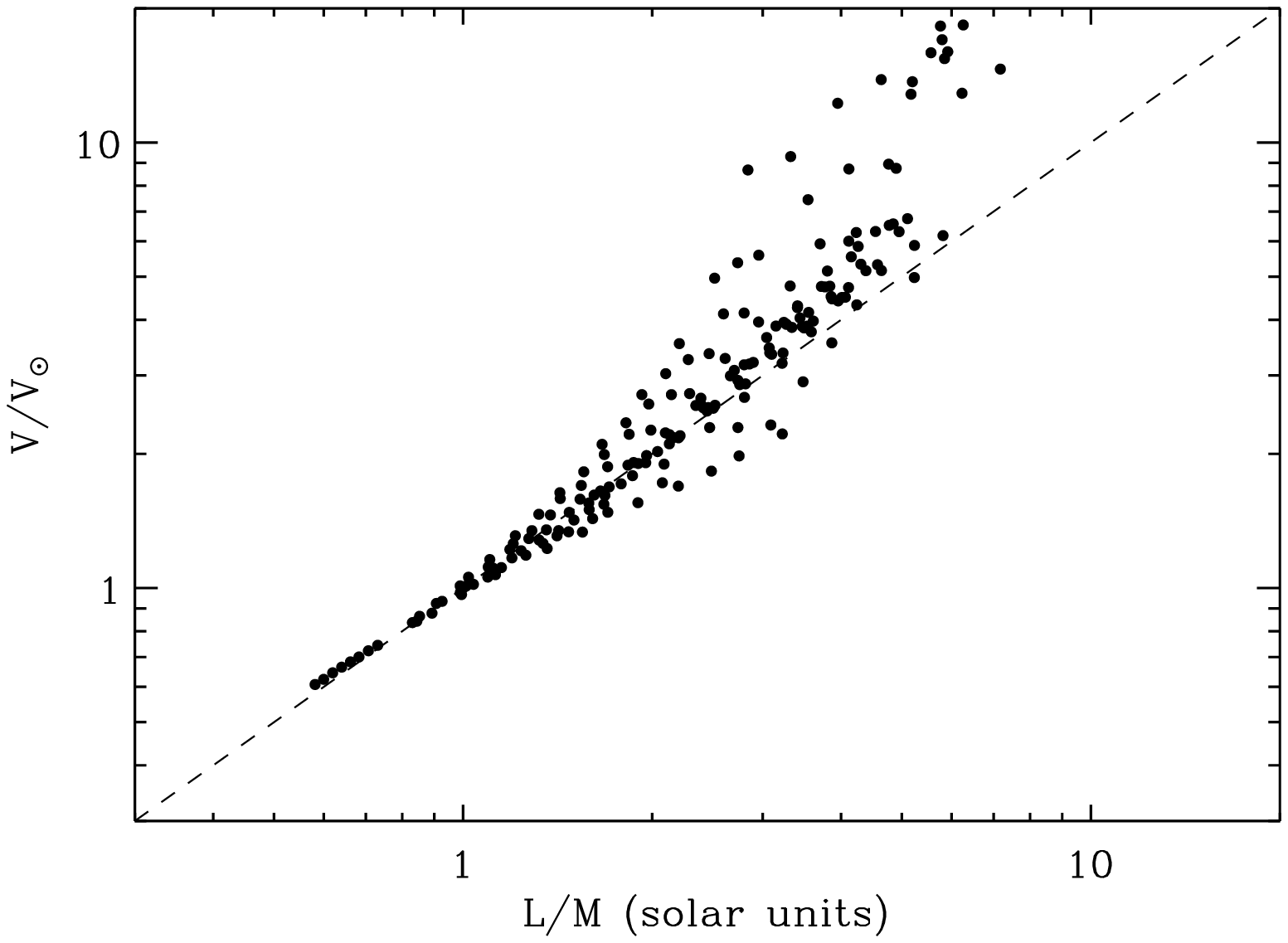}
\endfig

The low-temperature extremities of the contours indicate where the model 
hydrogen mass fractions $X_{\rm c}$ reach $10^{-6}$ in the core; no
calculations were carried out at lower temperatures. 
The amplitudes increase steeply with luminosity, particularly for stars with
mass $M\ga 1.4\,M_\odot$, owing to the increase in the convective velocities
with $M$. The largest amplitudes are predicted for a 1.6\,$M_\odot$ 
model of spectral type F2, which has a maximum velocity amplitude of 
$\sim 15$ times larger than that found for the Sun. For this model the 
turbulent Mach number is also predicted to be largest (see Fig.~12).
For more massive stars the computations predict overstable modes (see
section 9). Amplitudes of such overstable modes are limited by
nonlinear processes and can therefore not be estimated with the linear
computations adopted in this paper; their values could be much larger
than the amplitude values of the stable stochastically driven modes 
considered here.

\begfig 12.35 cm
\figure{16}
{Luminosity amplitudes (computed at outermost meshpoint)
versus effective temperature for models with constant
luminosity. The results are displayed for different mixing-length parameters
$\ac$ (top) and metallicities (bottom). 
The computations assumed the nonlocal convection parameters of Fig~4.
The line styles are as defined in Fig.~14.
{\bf Top:} the thick curves display the results for models 
computed with $\ac=1.8$ and the thin curves depict the amplitudes
obtained with $\ac=2.0$. In both model sequences the
value for the metallicity $Z$ was chosen to be $0.02$.
{\bf Bottom:} the thick curves depict the amplitudes from model calculations
using $Z=0.04$ and the thin curves for $Z=0.02$, assuming $\ac=2.0$.}
\includegraphics{\figdir/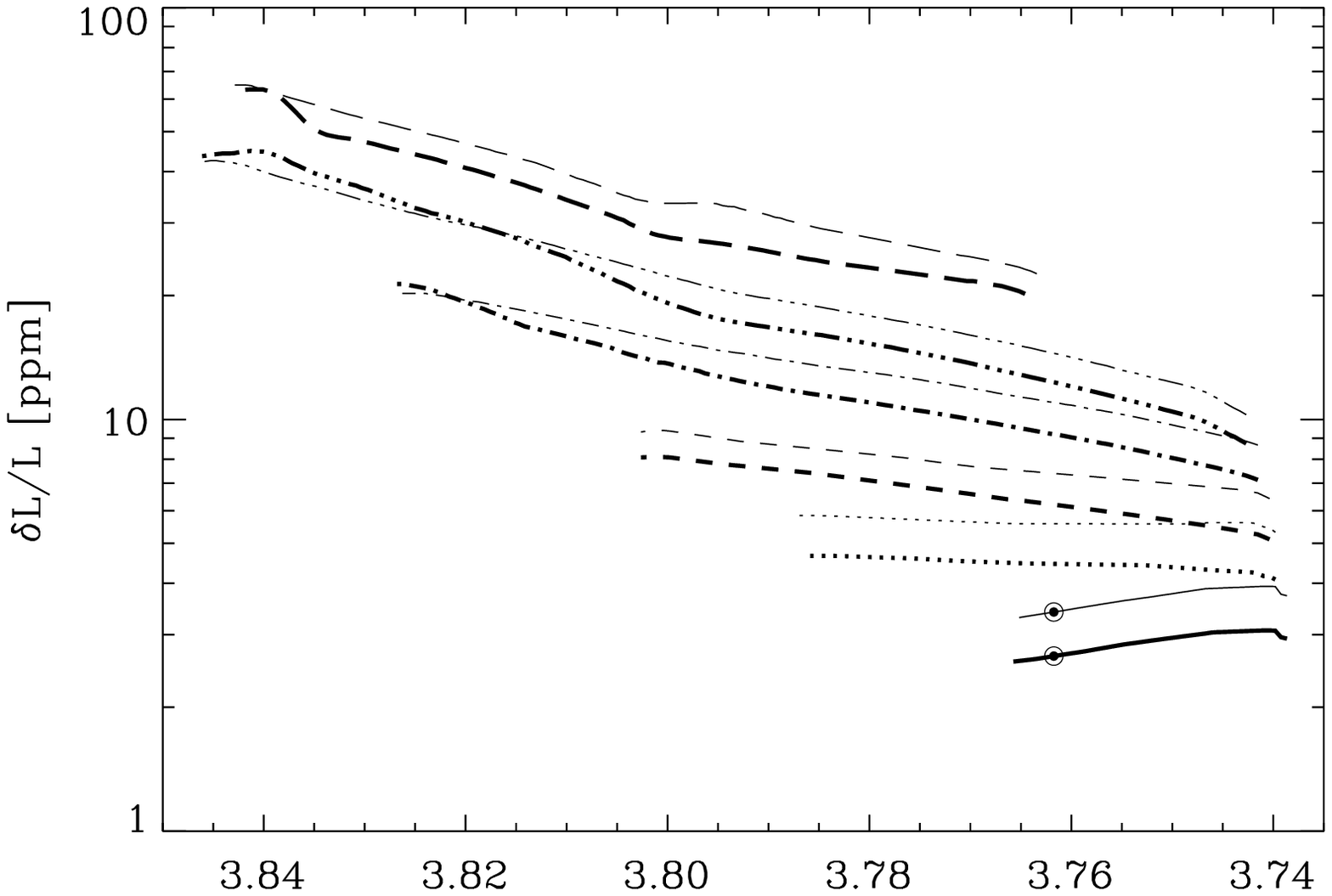}
\includegraphics{\figdir/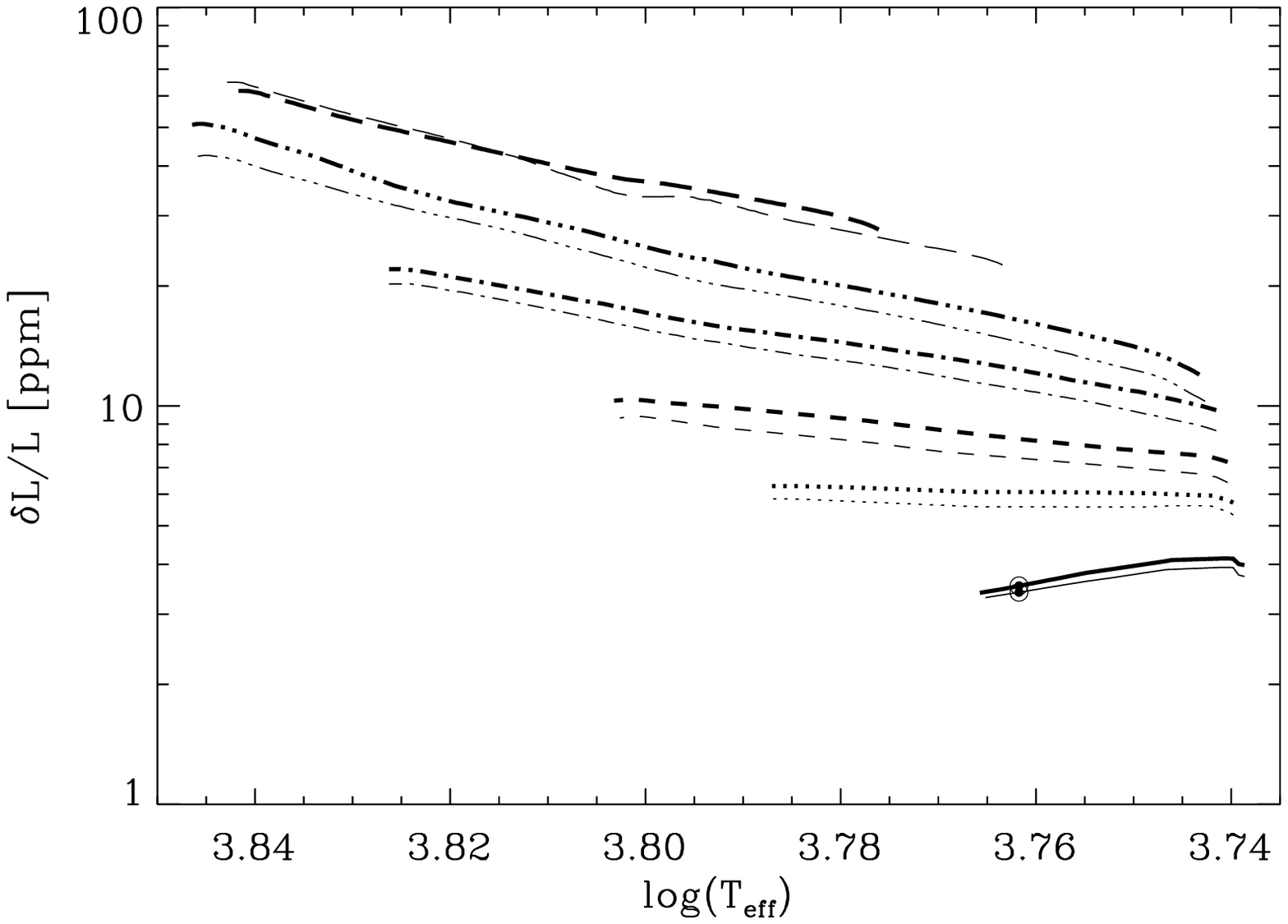}
\endfig

For stars with $L\la 2\,L_\odot$, the velocity amplitudes of stochastically 
excited modes depend strongly on the model luminosity and are only weakly 
dependent on the effective temperature. The same trend is also
seen in Fig.~14 for the luminosity amplitudes (top panel). 
In the lower panel of Fig.~14 the ratios between luminosity and velocity 
amplitudes, $\Delta L/\Delta V_{\rm s}$, are displayed, where the luminosity
fluctuations, $\Delta L$, are computed at the outermost meshpoint of the 
models. At the photospheric level ($h=0$\,km) the amplitude ratios 
appear to be quite insensitive to luminosity, and
depend mainly on effective temperature. 

Based on the model results of 
Christensen-Dalsgaard \& Frandsen (1983b), Kjeldsen \& Bedding (1995)
proposed a scaling relationship for solar-type velocity amplitudes
as a function of parameters used in stellar-evolution theory.
In particular, they proposed the scaling law
$$
  {V\over V_\odot}\sim\Bigl({L/L_\odot\over M/M_\odot}\Bigr)^s
  \,,\eqno\autnum
$$
with $s=1$, suggesting that the velocity amplitudes scale directly with the
light-to-mass ratio $L/M$ of the star. In Fig.~15 the velocity amplitudes
versus the light-to-mass ratio are displayed at a height $h=200$\,km 
above the photosphere for model calculations assuming the convection 
parameters of Fig.~4. There is a fair
agreement between the computed amplitudes (filled circles) and 
Kjeldsen \& Bedding's proposed relation (dashed line) for $L/M\la 3$.
For higher values of $L/M$ the estimated amplitudes are predicted to be 
larger than Kjeldsen \& Bedding's linear relation, particularly for models 
with masses $M/M_\odot\ga 1.4$. Moreover, for these models the computed
amplitudes become progressively more dependent on the model's effective 
temperature and less dependent on $L/M$ as they evolve along their 
evolutionary tracks (see also Fig.~13). 
Applying a linear polynomial fit to the estimated amplitudes in Fig.~15 
suggests for the exponent $s$ in the scaling law (9) a value of $1.29$. 
At the photospheric height the computations suggest a value of $s=1.47$.

We should point out, however, that the convective velocities
found in our models are large. As already indicated in Fig.~12,
the turbulent Mach number $M_{\rm t}$ becomes relatively large for models
with $M/M_\odot\ga 1.4$. Relative to a local convection model, the nonlocal 
formulation used here reduces the convective velocities, although they 
still remain large. This reduction results in part from the averaging of the 
superadiabatic temperature gradient over the eddies, which spreads the 
influence of this gradient's sharp peak in the hydrogen ionization zone 
over a larger region.

\titleb{Dependence on mixing length and metallicity}
The dependence of the luminosity amplitudes on mixing-length parameter 
$\ac$ and metallicity $Z$ is illustrated in Fig.~16 over a range of effective 
temperature for models with constant luminosity.
The dependence of the velocity amplitudes are illustrated in Fig.~17 for
two evolving models with mass $1.0M_\odot$ and $1.3M_\odot$. 

\begfig 12.35 cm
\figure{17}
{Velocity amplitudes for an evolving $1.0M_\odot$ and $1.3M_\odot$ star 
versus the model's effective temperature. The computations assumed the 
nonlocal convection parameters of Fig.~4 and results are displayed at a 
photospheric level $h=200$\,km.
{\bf Top:} amplitudes are depicted for three values of of the
mixing-length parameter $\ac$ assuming $Z=0.02$ in the computations.
{\bf Bottom:} results are plotted for three values of metallicity $Z$
assuming $\ac=2.0$ in the model calculations.}
\includegraphics{\figdir/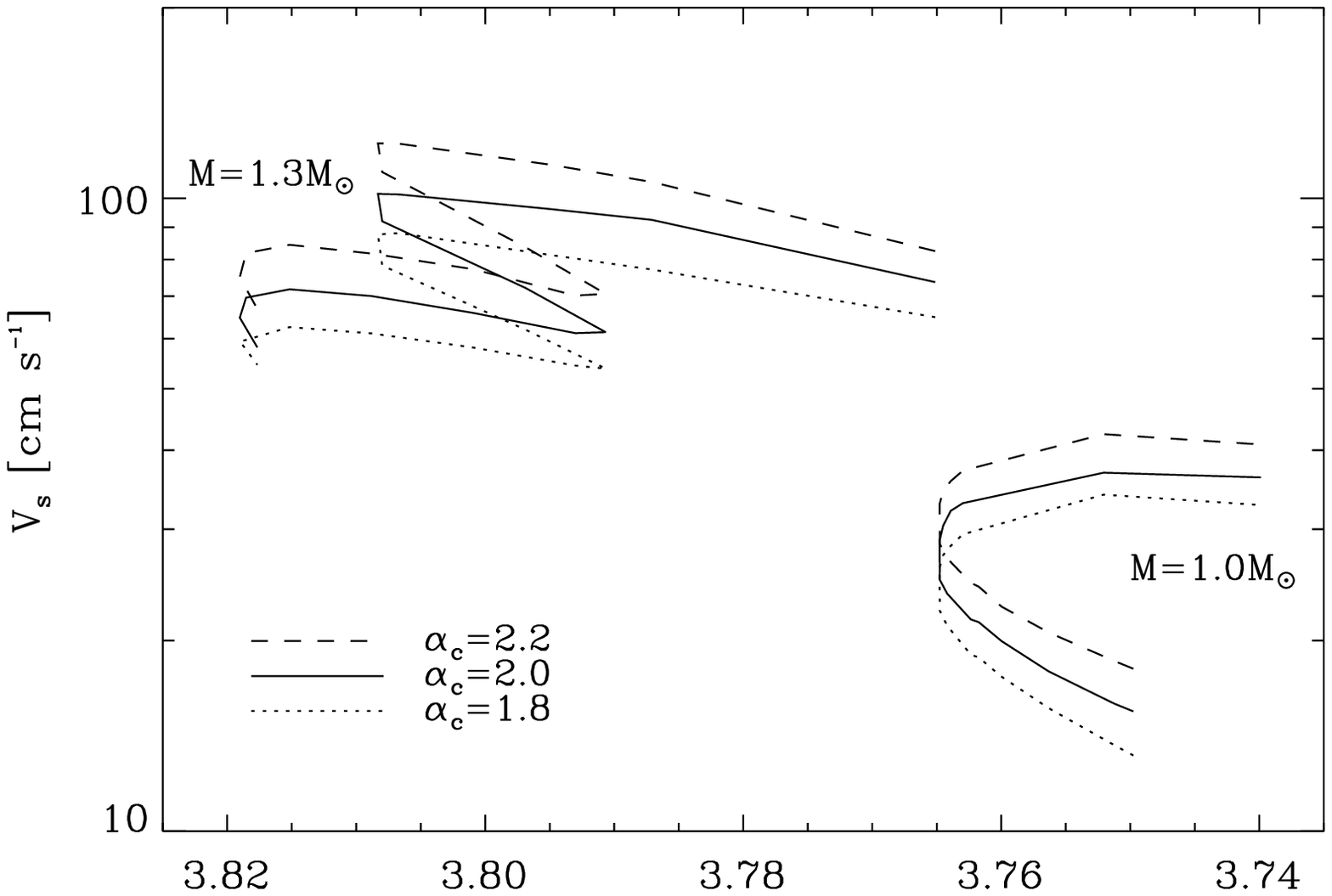}
\includegraphics{\figdir/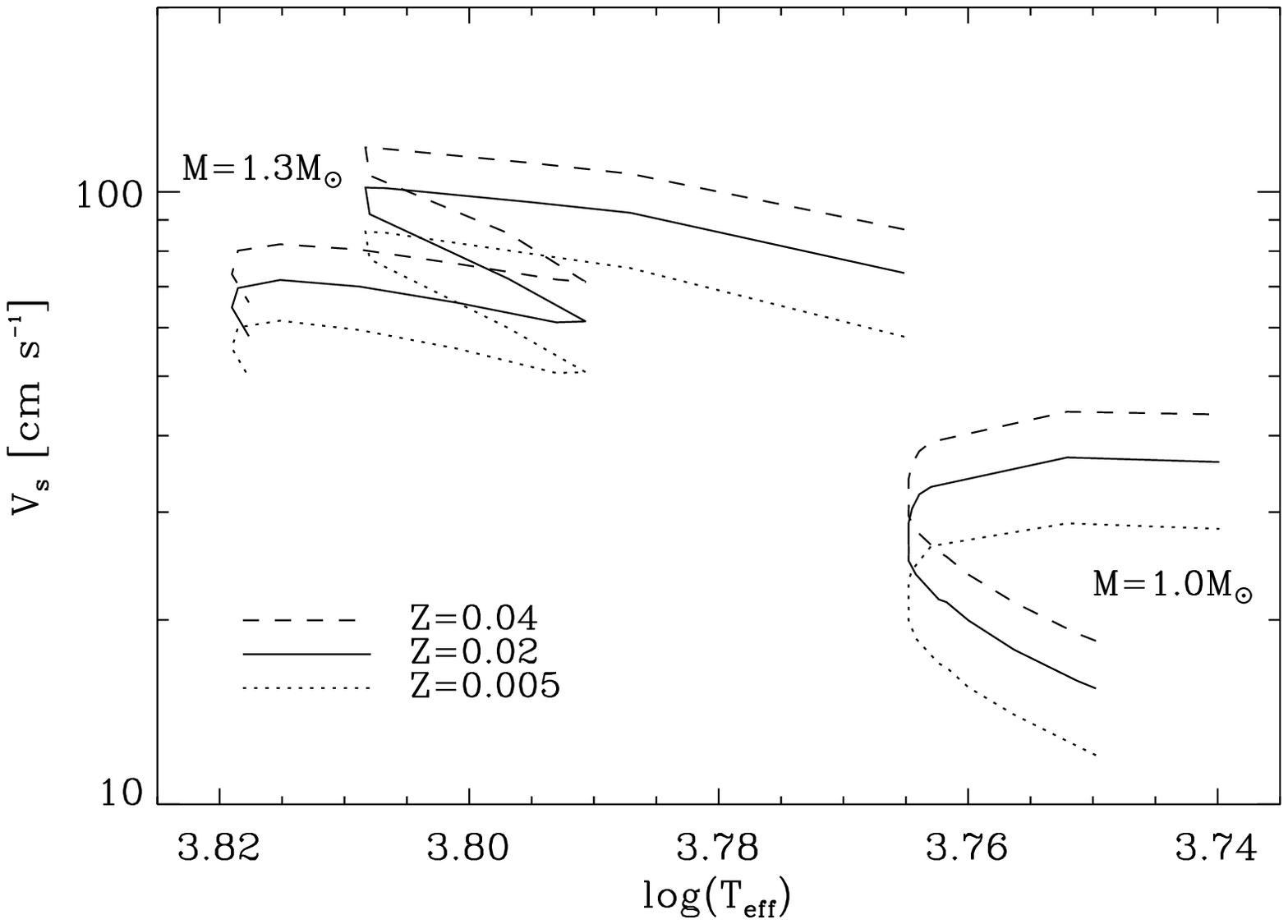}
\endfig

Increasing the
mixing-length parameter results in an increase of both the luminosity and
velocity amplitudes (top panels). This comes about because increasing $\ac$
results in higher convective velocities $u$ and thus in a larger acoustic
generation rate $P_{\rm Q}$ (see Eq.\ 8). 
Moreover, it appears that
the amplitudes become less dependent upon $\ac$ with increasing luminosity,
which might be explained by the decrease in convective efficacy with model
mass.

A similar behaviour of the amplitudes is obtained when the metallicity 
is increased (bottom panels in Figs~16 and 17). 
A larger value for the heavy-element abundance $Z$ results in a higher 
opacity $\kappa$ and consequently in a larger convective heat flux in the 
upper boundary layer of the convection zone. Therefore the turbulent 
Mach number $M_{\rm t}$ becomes larger, and thus also do the amplitudes.

\titlea{Overstable modes}
For stars with log\,$T_{\rm eff}\ga 3.85$ lying more or less in the 
$\delta$\, Scuti instability strip, the model calculations predict modes 
to be overstable, irrespectively of adjustments to the convection parameters
$a$ and $b$. The $\delta$\,Scuti stars are variables with
spectral types A and F in the lower part of the classical
Cepheid instability strip, which are in the very interesting
evolutionary phase of the main sequence near the end of central 
hydrogen burning. It was first shown by Zhevakin (1953) and 
Cox \& Whitney (1958) that the excitation mechanism in Cepheids, 
which are core-helium-burning radial pulsators having large amplitudes 
of the fundamental mode (and in a few cases also the first overtone),
is plausibly due to the opacity mechanism acting in the HeII ionization zone
(see also Baker \& Kippenhahn 1962).
The same mechanism is believed to be responsible for the excitation
in $\delta$\,Scuti stars (e.g. Dziembowski 1995 and references therein).
The oscillation specta of many $\delta$\,Scuti stars, however,
are far more complex, involving both radial and nonradial modes with 
low amplitudes, lying often in a narrow frequency
range. This complicates mode identification substantially
(e.g. Mangeney\etal\ 1991).
The cooler $\delta$\,Scuti stars have substantial outer convection zones.
Thus in these 
layers the pulsationally induced fluctuations of the turbulent fluxes 
may become important for the selection mechanism of modes with observable 
amplitudes.

The theoretically predicted order $n$ of unstable
p modes in sequences of evolving models of $\delta$\,Scuti stars 
are depicted by different symbols
in Fig.~13 (e.g. circles indicate the location of models 
in the HR diagram for which the radial fundamental mode was 
found to be overstable). The models have only a few excited 
modes lying in a narrow frequency interval, and some of them 
display radial orders in a nonconsecutive sequence
(however, see also Houdek \& Gough 1998; Michel\etal\ 1999).
Moreover, with increasing effective temperature the overstable modes
shift to higher frequencies.
The blue edge of the instability domain is found to shift to higher
effective temperatures with increasing order; this result is consistent 
with previous model calculations (e.g. Stellingwerf 1979, 1980;
Dziembowski 1995) mainly because for these models convection is
unimportant. Through the inclusion of the
turbulent flux perturbations in the stability analyses the
computations predict well defined red edges, a result which was
previously reported by Baker \& Gough (1979) for RR Lyrae stars.
In particular, the fluctuating Reynolds stress $\delta p_{\rm t}$ 
is found to be the decisive contributor to the damping rates and 
thus for the return to stability at the red edge for low-order 
modes in $\delta$ Scuti stars (Houdek 1997). 
Only with the inclusion of $\delta p_{\rm t}$ in the computations are
all modes found to be stable for models with effective temperatures 
satisfying log\,$T_{\rm eff}\la\,3.85$.

\titlea{Conclusion}
It is evident that one of the greatest deficiencies in modelling 
oscillations in stars with surface convection zones is the lack of 
a proper theory of convection in a pulsating environment.
Although several attempts have been made in recent years to
address this problem (for a review, see Baker 1987) none of the proposed
prescriptions are anything more than phenomenological.
Impressive progress has been made on
hydrodynamical simulations of convection, including also the
interaction with pulsations (e.g. Stein \& Nordlund 1991;
Bogdan, Cattaneo \& Malagoli 1993; Nordlund \& Stein 1998;
Stein \& Nordlund 1998).
In particular, the work by Stein and Nordlund, including a
realistic treatment of the physics of the outer parts of the
convection zone, has confirmed the earlier
conclusion that the solar oscillations
are likely to be intrinsically stable 
(e.g. Gough 1980; Kumar \& Goldreich 1989; Balmforth 1992a);
also, the simulations yielded estimates,
similar to the observationally determined values,
of the energy input
to the modes from the stochastic driving by convection.
However, such simulations are evidently extremely time consuming
and have so far been made in sufficient detail only for solar parameters.
Here we have estimated the pulsational properties of main-sequence
stars over a broad range of parameters, by means of 
a time-dependent non-local version of mixing-length theory.

Perhaps the most important conclusion drawn from this
survey is that, as in the case of the Sun, oscillations in solar-like stars
are intrinsically damped and stochastically driven by convection.
We note that the issue may still not be entirely settled, however.
In particular, Cheng \& Xiong (1997) reported calculations using 
Xiong's (1989) nonlocal formulation of mixing-length theory which 
predict overstable solar oscillations. In fact, their computations 
suggest that the momentum flux perturbations destabilize all p modes, 
in complete disagreement to the results reported here.
This discrepancy evidently deserves investigation.

\acknow{We are grateful to W. Chaplin for providing us with the latest
BiSON data. This work was supported in part by Danmarks Grundforskningsfond 
through the establishment of the Theoretical Astrophysics Center and by
the Particle Physics and Astronomy Research Council of the UK. 
G.H. acknowledges travel support by the Austrian FWF (project S7303-AST).
}

\begref{References}
\ref Antia H.M., Chitre S.M., Gough D.O., 1988,
in: {\rm Proc. IAU Symposium No 123, Advances in helio- and asteroseismology},
Christensen-Dalsgaard J., Frandsen S. (eds).
Reidel, Dordrecht, p. 371 
\ref Appourchaux T. and the VIRGO team, 1998, in: Structure and dynamics of the
     interior of the Sun and Sun-like stars; Proc. SOHO 6/GONG 98 Workshop,
     Korzennik, S.G., Wilson, A. (eds). ESA SP-418, Noordwijk, p. 99
\ref Auer L.H., Mihalas D., 1970, MNRAS 149, 65 
\ref Baker N.H., 1987, in: Physical Processes in Comets,
     Stars and Active Galaxies, Hillebrandt, W., Meyer-Hofmeister, E.,
     Thomas, H.-C. (eds). Springer-Verlag, New York, p.~105 
\ref Baker N.H., Kippenhahn R., 1962, Zs. f. Ap. 54, 114 
\ref Baker N.H., Kippenhahn R., 1965, ApJ 142, 868 
\ref Baker N.H., Gough D.O., 1979 ApJ, 234,232 
\ref Baker N.H., Moore D.W., Spiegel E.A., 1971,
     Q. J. Mech. Appl. Math. 24, 391 
\ref Balmforth N.J., 1992a, MNRAS 255, 603 
\ref Balmforth N.J., 1992b, MNRAS 255, 639 
\ref Balmforth N.J., Gough D.O., 1990a, SPh 128, 161 
\ref Balmforth N.J., Gough D.O., 1990b, ApJ 362, 256 
\ref Batchelor G.K., 1956, The theory of homogeneous turbulence,
     Cambridge University Press 
\ref Bedding T.R., Kjeldsen H., Reetz J, Barbuy B., 1996, MNRAS 280, 1155 
\ref Bogdan T.J., Cattaneo F., Malagoli A., 1993, ApJ 407, 316 
\ref B\"ohm-Vitense E., 1958, Zs. F. Ap. 46, 108 
\ref Brown T.M., Gilliland R.L., 1990, ApJ 350, 839 
\ref Brown T.M., Gilliland R.L., Noyes R.W., Ramsey L.W., 1991, ApJ 368, 599 
\ref Canuto V.M., Christensen-Dalsgaard J., 1998, Ann. Rev. Fluid Mech. 30,
     167 
\ref Catala C. and the COROT team, 1995, in:
     Proc. GONG'94 Helio- and Astero-seismology from Earth and Space,
     Ulrich, R.K., Rhodes Jr, E.J., D\"appen, W. (eds). PASPC 76,
     San Francisco, p.~426 
\ref Chang H-Y., Gough D.O., Sekii T., 1997,
     in: Proc. IAU Symp. 181: Sounding Solar and Stellar Interiors,
     Schmider, F.-X., Provost, J. (eds). Nice Observatory, 
     Poster Volume, p. 13 
\ref Chaplin W.J., Elsworth Y., Isaak G.R., Lines R., McLeod C.P.,
     Miller B.A., New R., 1998, MNRAS 298, 7 
\ref Cheng Q.L., Xiong D.R., 1997, A\&A 319, 981 
\ref Christensen-Dalsgaard J., 1982, MNRAS 199, 735 
\ref Christensen-Dalsgaard J., 1993, in: Inside the Stars,
     Weiss, W.W., Baglin, A. (eds). PASPC~40, San Francisco, p. 483 
\ref Christensen-Dalsgaard J., Frandsen S., 1983a, SPh 82, 165  
\ref Christensen-Dalsgaard J., Frandsen S., 1983b, SPh 82, 469  
\ref Christensen-Dalsgaard J., Gough D.O., 1982, MNRAS 198, 141 
\ref Christensen-Dalsgaard J., Gough D.O., Libbrecht K.G., 1989, ApJ 341, L103 
\ref Cox J.P., Whitney C., 1958, ApJ 127, 561 
\ref Dziembowski W.A., 1995,
     in: Proc. GONG'94: Helio- and Astero-seismology from Earth and Space,
     Ulrich, R.K., Rhodes Jr, E.J., D\"appen, W. (eds). PASPC~76,
     San Francisco, p. 586 
\ref Elsworth Y., Howe R., Isaak G.R., Mcleod C.P.,
     Miller B.A., New R., Speake, C.C., Wheeler S.J., 1993, MNRAS 265, 888 
\ref Eggleton P., Faulkner J., Flannery B.P., 1973, A\&A 23, 325 
\ref Frandsen S., 1992, in: Inside the Stars,
     Weiss W.W., Baglin A. (eds). PASPC~40, San Francisco, p. 679 
\ref Fr\"ohlich C. and the VIRGO team, 1995, SPh 162, 101 
\ref Gabriel M., 1998, A\&A 330, 359 
\ref Gabriel A.H. and the GOLF team, 1991, Adv. Space Res., vol. 11,
     No. 4, 103 
\ref Gilliland R.L., 1995,
     in: Proc. GONG'94: Helio- and Astero-seismology from Earth and Space,
     Ulrich, R.K., Rhodes Jr, E.J., D\"appen, W. (eds). PASPC~76,
     San Francisco, p.~578 
\ref Gilliland R.L., Brown T.M., Kjeldsen H., McCarthy J.K., Peri M.L.,
     Belmonte J.A., Vidal I., Cram L.E., Palmer J., Frandsen S.,
     Parthasarathy M., Petro L., Schneider H., Stetson P.B., Weiss W.W., 1993,
     AJ 106, 2441 
\ref Goldreich P., Keeley D.A., 1977a, ApJ 211, 934 
\ref Goldreich P., Keeley D.A., 1977b, ApJ 212, 243 
\ref Goldreich P., Kumar P., 1990,
{ApJ} {\rm 363}, 694 
\ref Goldreich P., Kumar P., 1991, ApJ 374, 366 
\ref Goldreich P., Murray N., 1994, ApJ 424, 480 
\ref Goldreich P., Murray N., Kumar P., 1994, ApJ 423, 466 
\ref Gonczi G., Osaki Y., 1980, A\&A 84, 304 
\ref Goode P.R., Strous L.H., 1996, Bull. Astr. Soc. India 24, 223 
\ref Goode P.R., Gough D.O., Kosovichev A., 1992, ApJ 387, 707 
\ref Gough D.O., 1965, in: Geophysical Fluid Dynamics II,
     Woods Hole Oceanographic Institution, p. 49
\ref Gough D.O., 1976, in: Problems of stellar convection,
     Spiegel, E., Zahn, J.-P. (eds). Springer-Verlag, Berlin, p. 15 
\ref Gough D.O., 1977, ApJ 214, 196 
\ref Gough D.O., 1980, in: Nonradial and Nonlinear Stellar Pulsation,
     Hill, H.A., Dziembowski, W.A. (eds). Springer-Verlag, Berlin, p. 273 
\ref Gough D.O., 1995,
     in: Proc. GONG'94: Helio- and Astero-seismology from Earth and Space,
     Ulrich, R.K., Rhodes Jr, E.J., D\"appen, W. (eds). PASPC~76,
     San Francisco, p. 331 
\ref Gough D.O., 1997,
     in: Proc. IAU Symp. 181: Sounding Solar and Stellar Interiors,
     Schmider, F.-X., Provost, J. (eds). Nice Observatory, p. 397 
\ref Gough D.O., Weiss N.O., 1976, MNRAS 176, 589 
\ref Gough D.O., Spiegel E.A., Toomre J., 1974, 
     Lecture Notes in Physics~35, Richtmeyer, R. (ed.). 
     Springer-Verlag, Heidelberg 
\ref Grec G., Fossat E., Pomerantz M., 1983, SPh 82, 55 
\ref Harvey J.W.\etal , 1996, Science 272, 1284 
\ref Henyey L., Vardya M.S., Bodenheimer P., 1965, ApJ 142, 841 
\ref Hill F.\etal , 1996, Science 272, 1292 
\ref Houdek G., 1996, Ph.D. Thesis, Universit\"at Wien 
\ref Houdek G., 1997, in: Proc. IAU Symp. 181: Sounding Solar and Stellar 
     Interiors, Schmider, F.-X., Provost, J. (eds). Nice Observatory, 
     Poster Volume, p. 227 
\ref Houdek G., Rogl J., 1996, Bull. Astr. Soc. India 24, 317 
\ref Houdek G., Gough D.O., 1998, in: Structure and dynamics of the
     interior of the Sun and Sun-like stars; Proc. SOHO 6/GONG 98 Workshop,
     Korzennik, S.G., Wilson, A. (eds). ESA SP-418, Noordwijk, p. 479
\ref Houdek G., Rogl J., Balmforth N., Christensen-Dalsgaard J., 1995,
     in: Proc. GONG'94: Helio- and Astero-seismology from Earth and Space,
     Ulrich, R.K., Rhodes Jr, E.J., D\"appen, W. (eds). PASPC~76,
     San Francisco, p. 641 
\ref Iglesias C.A., Rogers F.J., 1996, ApJ 464, 943 
\ref Kennelly E.J., 1995, in: Proc. GONG'94: Helio- and Astero-seismology 
     from Earth and Space, Ulrich, R.K., Rhodes Jr, E.J., D\"appen, W. (eds). 
     PASPC~76, San Francisco, p. 568 
\ref Jefferies S.M., Pall\'e P.L., van der Raay H.B., R\'egulo C.,
     Roca Cort\'es T., 1988, Nature 333, 646 
\ref Kjeldsen H., Bedding T.R., 1995, A\&A 293, 87 
\ref Kjeldsen H., Bedding T.R., 1998, in: Proc. First MONS Workshop,
     Kjeldsen, H., Bedding, T.R. (eds). Aarhus Universitet, Denmark, p. 1
\ref Kjeldsen H., Bedding T.R., Viskum M., Frandsen S., 1995,
{AJ} {\rm 109}, 1313 
\ref Kumar P., Goldreich P., 1989, ApJ 103, 331 
\ref Kurucz R.L., 1991, in: Stellar Atmospheres: Beyond Classical Models,
     Crivellari, L., Hubeny, I., Hummer, D.G. (eds). Kluwer, Dordrecht, p. 441 
\ref Lamb H., 1909, Proc. London Math. Soc. 7, 122 
\ref Libbrecht K.G., 1988, ApJ, 334, 510 
\ref Libbrecht K.G., Woodard M.F., 1991, Science 253, 152 
\ref Libbrecht K.G., Popp B.D., Kaufman J.M., Penn M.J., 1986,
     Nature 323, 235 
\ref Mangeney A., D\"appen W., Praderie F., Belmonte J.A., 1991, A\&A 244, 351 
\ref Matthews J.M., 1998, in: Structure and dynamics of the
     interior of the Sun and Sun-like stars; Proc. SOHO 6/GONG 98 Workshop,
     Korzennik, S.G., Wilson, A. (eds). ESA SP-418, Noordwijk, p. 395
\ref Michel E., Hern\'andez M.M., Houdek G., Goupil M.J., Lebreton Y., 
     Hern\'andez F.P\'erez, Baglin A., Belmonte J.A., Soufi F., 1999, 
     A\&A 342, 153
\ref Moore, D.W., Spiegel E.A., 1964, ApJ 139, 48 
\ref Mosser B., Maillard J.P., M\'ekarnia D., Gay J., 1998, A\&A 340, 457
\ref Musielak Z.E., Rosner R., Stein R.F., Ulmschneider P., 1994,
     ApJ 423, 474 
\ref Nigam R., Kosovichev A.G., Scherrer P.H., Schou J., 1998, ApJ 495L, 115N 
\ref Nordlund {\AA}., Stein R.F., 1998, in: Proc. IAU Symp. 185: 
     New eyes to see inside the Sun and stars, Deubner, F.-L., 
     Christensen-Dalsgaard, J., Kurtz, D.W. (eds). Kluwer, Dordrecht, p. 199 
\ref Osaki Y., 1990, in: Progress of Seismology of the Sun and Stars,
     Osaki, Y., Shibahashi, H. (eds). Springer-Verlag, Berlin, p. 145 
\ref Rast M.P., Bogdan T.J., 1998,
{ApJ} {\rm 496}, 527 
\ref Rosenthal C.S., Christensen-Dalsgaard J., Houdek G.,
     Monteiro M.J.P.F.G., Nordlund \AA., Trampedach R., 1995,
     in: Proc. 4th SOHO Workshop: Helioseismology,
     Hoeksema, J.T., Domingo, V., Fleck, B., Battrick, B. (eds).
     ESA SP-376, vol.2, ESTEC, Noordwijk, p. 459
\ref Roxburgh I.W., Vorontsov S.V., 1997,
{MNRAS} {\rm 292}, L33 
\ref Scherrer P.H., Hoeksema J.T., Bush R.I., 1991, Adv. Space Res.,
     vol. 11, No. 4, 113 
\ref Schrijver C.J., Jim\'enez A., D\"appen W., 1991, A\&A 251, 655 
\ref Spiegel E., 1962, J.~Geophys.~Res.~67, 3063 
\ref Spiegel E., 1963, ApJ 138, 216 
\ref Stein R.F., 1967, SPh 2, 385 
\ref Stein R.F., Nordlund \AA., 1991, in: Challenges to Theories of the
     Structure of Moderate Mass Stars, Gough, D.O., Toomre, J. (eds).
     Springer-Verlag, Heidelberg, p. 195 
\ref Stein R.F., Nordlund {\AA}., 1998, in: Structure and dynamics of the
     interior of the Sun and Sun-like stars; Proc. SOHO 6/GONG 98 Workshop,
     Korzennik S.G., Wilson A. (eds). ESA SP-418, Noordwijk, p. 693
\ref Stellingwerf R.F., 1979, ApJ 227, 935 
\ref Stellingwerf R.F., 1980, in: Nonradial and Nonlinear Stellar Pulsation,
     Hill, H.A., Dziembowski, W.A. (eds). Springer-Verlag, Berlin, p. 50 
\ref Tooth P.D., Gough D.O., 1989, in: Seismology
     of the Sun and Sun-like Stars, Domingo, V., Rolfe, E. (eds).
     ESA SP-286, Noordwijk, p. 463 
\ref Toutain T., Fr\"ohlich C., 1992, A\&A 257, 287 
\ref Ulrich R.K., 1970a, ApJ 162, 993 
\ref Ulrich R.K., 1970b, Ap\&SS 7, 71 
\ref Unno W., 1964, Trans. Int. astr. Un. XII(B), 555 
\ref Unno W., 1967, PASJ 19, 140 
\ref Unno W., Kato S., 1962, PASJ 14, 416 
\ref Unno W., Spiegel E.A., 1966, PASJ 18, 85 
\ref Vernazza J.E., Avrett E.H., Loeser R., 1981, ApJS 45, 635 
\ref Xiong D., 1989, A\&A 254, 362 
\ref Zhevakin S.A., 1953, {\rm Astron. Zh.} 30, 161 
\endref
\bye